\documentclass[12pt,onecolumn,
tightenlines,
superscriptaddress,nofootinbib,preprintnumbers]{revtex4-2}
\makeatletter
\def\l@subsubsection#1#2{}
\makeatother

\usepackage[utf8]{inputenc}
\newcommand{\mysize}{1.1}

\usepackage{csquotes}

\usepackage[dvipsnames]{xcolor}
\usepackage{xspace}
\usepackage{graphicx}
\usepackage{cancel} 
\usepackage{multirow}
\graphicspath{{./figures/}}
\usepackage[caption=false]{subfig}

\usepackage{tikz}
\usetikzlibrary{
  arrows.meta,
  calc,
  decorations.markings,
  fit,
  matrix,
  positioning,
  shapes
}
\tikzset{
  massless/.style={line width=0.2ex},
  masslesscollinear/.style={line width=0.3ex, teal, dotted},
  bigmassless/.style={line width=0.3ex},
  reduced/.style={line width=0.2ex, dashed, dash phase=3pt},
  dotdot/.style={line width=0.2ex, dashed, dash phase=2pt},
  loop arrow/.style={postaction={decorate, decoration={markings, mark=at position 0.65 with {\arrow{Stealth}}}}}
}
\newlength{\sideLength}

\newcommand{\UVdoubleTriangle}{
  \begin{tikzpicture}
    [baseline=-3.5ex, x=6ex, y=6ex]
      \setlength{\sideLength}{2cm} 

    \coordinate (IN) at (-1,0); 
    \coordinate (OUT) at (3,0); 
    \coordinate (A) at (0,0);
    \coordinate (B) at (2,0);
    \coordinate (C) at (1,1);
    \coordinate (D) at (1,-1);

     \draw (0.25,0.75) node{\(\ell_1\)};
     \draw (1.75,0.75) node{\(\ell_2\)};
      \draw[massless]  (A) --  (C) -- (B) -- (D) -- (A);
      \draw[massless] (C) -- (D);
      \draw[massless] (IN) node[left]{\(k\)} -- (A);
      \draw[massless] (B) -- (OUT);
  \end{tikzpicture}
}

\newcommand{\UVdoubleTriangleSubA}{
  \begin{tikzpicture}
    [baseline=0ex, x=6ex, y=6ex]
      \setlength{\sideLength}{2cm} 

    \coordinate (IN) at (-1,0); 
    \coordinate (OUT) at (3,0); 
    \coordinate (A) at (0,0);
    \coordinate (B) at (2,0);
    \coordinate (C) at (1,1);
    \coordinate (D) at (1,-1);

      \draw[massless]  (A) --  (C);
      \draw[reduced]  (C) -- (B);
      \draw[reduced]  (B) -- (D);
      \draw[massless]  (D) -- (A);
      \draw[reduced] (C) -- (D);
      \draw[massless] (IN) -- (A);
      \draw[massless] (B) -- (OUT);
  \end{tikzpicture}
}

\newcommand{\UVdoubleTriangleSubB}{
  \begin{tikzpicture}
    [baseline=0ex, x=6ex, y=6ex]
      \setlength{\sideLength}{2cm} 

    \coordinate (IN) at (-1,0); 
    \coordinate (OUT) at (3,0); 
    \coordinate (A) at (0,0);
    \coordinate (B) at (2,0);
    \coordinate (C) at (1,1);
    \coordinate (D) at (1,-1);

      \draw[reduced]  (A) --  (C);
      \draw[massless]  (C) -- (B);
      \draw[massless]  (B) -- (D);
      \draw[reduced]  (D) -- (A);
      \draw[reduced] (C) -- (D);
      \draw[massless] (IN) -- (A);
      \draw[massless] (B) -- (OUT);
  \end{tikzpicture}
}

\newcommand{\UVdoubleTriangleSubC}{
  \begin{tikzpicture}
    [baseline=0ex, x=6ex, y=6ex]
      \setlength{\sideLength}{2cm} 

    \coordinate (IN) at (-1,0); 
    \coordinate (OUT) at (3,0); 
    \coordinate (A) at (0,0);
    \coordinate (B) at (2,0);
    \coordinate (C) at (1,1);
    \coordinate (D) at (1,-1);

      \draw[reduced]  (A) --  (C);
      \draw[reduced]  (C) -- (B);
      \draw[reduced]  (B) -- (D);
      \draw[reduced]  (D) -- (A);
      \draw[massless] (C) -- (D);
      \draw[massless] (IN) -- (A);
      \draw[massless] (B) -- (OUT);
  \end{tikzpicture}
}

\newcommand{\SBox}{
  \begin{tikzpicture}[scale=0.6]
    [baseline=-3.5ex, x=6ex, y=6ex]
    \coordinate (v1) at (0,0);
    \coordinate (v2) at (4,0);
    \coordinate (v3) at (4,-4);
    \coordinate (v4) at (0,-4);

    \draw[massless](v1) -- (v2) node[midway, above]{\large \(q_3,\alpha_3\)};
    \draw[massless](v2) -- (v3) node[midway, right]{\large\(q_4,\alpha_4\)};
    \draw[massless](v3) -- (v4) node[midway, below]{\large\(q_1,\alpha_1\)};
    \draw[massless](v4) -- (v1) node[midway, left]{\large\(q_2,\alpha_2\)};

    \draw[massless] (v4) -- +(-135:1.6) node[left]{\large\(k_1\)};
    \draw[massless] (v1) -- +(135:1.6) node[left]{\large\(k_2\)};
    \draw[massless] (v2) -- +(45:1.6) node[right]{\large\(k_3\)};
    \draw[massless] (v3) -- +(-45:1.6) node[right]{\large\(k_4\)};
  \end{tikzpicture}
}

\newcommand{\Pentagon}{
  \begin{tikzpicture}[scale=0.7]
    [baseline=-3.5ex, x=6ex, y=6ex]
      \setlength{\sideLength}{2cm} 
    
      \coordinate (A) at (90:\sideLength);
      \coordinate (B) at (18:\sideLength);
      \coordinate (C) at (-54:\sideLength);
      \coordinate (D) at (234:\sideLength);
      \coordinate (E) at (162:\sideLength);
    
     \draw[massless] (A) -- (B) -- (C) -- (D) -- (E) -- cycle;
     \draw[massless] (A) -- +(90:0.8cm) node[left]{\large\(k_1\)};
     \draw[massless] (B) -- +(18:0.8cm) node[right]{\large\(k_2\)};
     \draw[massless] (C) -- +(-54:0.8cm) node[right]{\large\(k_3\)};
     \draw[massless] (E) -- +(162:0.8cm) node[left]{\large\(k_5\)};
     \draw[massless] (D) -- +(234:0.8cm) node[left]{\large\(k_4\)};

     \draw[massless, loop arrow] (E) -- (A) node[midway, above]{\large\(\ell\)};
  \end{tikzpicture}
}

\newcommand{\DBox}{
  \begin{tikzpicture}
    [baseline=
    0ex, x=6ex, y=6ex]
    \coordinate (v1) at (0,0);
    \coordinate (v2) at (1,0);
    \coordinate (v3) at (2,0);
    \coordinate (v4) at (0,-1);
    \coordinate (v5) at (1,-1);
    \coordinate (v6) at (2,-1);

    \draw[massless] (v4) -- (v1) -- (v2) -- (v3) -- (v6);
    \draw[massless] (v2) -- (v5);
    \draw[massless, loop arrow] (v5) -- (v4) node[midway, below]{\(\ell_1\)};
    \draw[massless, loop arrow] (v6) -- (v5) node[midway, below]{\(\ell_2\)};

    \draw[massless] (v4) -- +(-135:0.4) node[left]{\(k_1\)};
    \draw[massless] (v1) -- +(135:0.4) node[left]{\(k_2\)};
    \draw[massless] (v3) -- +(45:0.4) node[right]{\(k_3\)};
    \draw[massless] (v6) -- +(-45:0.4) node[right]{\(k_4\)};
  \end{tikzpicture}
}

\newcommand{\DBoxBeforeA}{
  \begin{tikzpicture}
    [baseline=-3.5ex, x=6ex, y=6ex]
    \coordinate (v1) at (0,0);
    \coordinate (v2) at (1,0);
    \coordinate (v3) at (2,0);
    \coordinate (v4) at (0,-1);
    \coordinate (v5) at (1,-1);
    \coordinate (v6) at (2,-1);

    \draw[massless] (v5) -- (v4) -- (v1) -- (v2);
    \draw[reduced] (v5) -- (v6) -- (v3) -- (v2) -- cycle;

    \draw[massless] (v4) -- +(-135:0.4);
    \draw[massless] (v1) -- +(135:0.4);
    \draw[massless] (v3) -- +(45:0.4);
    \draw[massless] (v6) -- +(-45:0.4);
  \end{tikzpicture}
}

\newcommand{\DBoxAfterA}{
  \begin{tikzpicture}
    [baseline=2.5ex, x=6ex, y=6ex]
    \coordinate (v1) at (0,0);
    \coordinate (v2) at (0,1);
    \coordinate (v3) at (1,0.5);

    \draw[massless] (v1) -- (v2) -- (v3) -- cycle;

    \draw[massless] (v1) -- +(-135:0.4);
    \draw[massless] (v2) -- +(135:0.4);
    \draw[massless] (v3) -- +(45:0.4);
    \draw[massless] (v3) -- +(-45:0.4);
  \end{tikzpicture}
}

\newcommand{\DBoxBeforeB}{
  \begin{tikzpicture}
    [baseline=-3.5ex, x=6ex, y=6ex]
    \coordinate (v1) at (0,0);
    \coordinate (v2) at (1,0);
    \coordinate (v3) at (2,0);
    \coordinate (v4) at (0,-1);
    \coordinate (v5) at (1,-1);
    \coordinate (v6) at (2,-1);

    \draw[massless] (v5) -- (v6) -- (v3) -- (v2);
    \draw[reduced] (v5) -- (v4) -- (v1) -- (v2) -- cycle;

    \draw[massless] (v4) -- +(-135:0.4);
    \draw[massless] (v1) -- +(135:0.4);
    \draw[massless] (v3) -- +(45:0.4);
    \draw[massless] (v6) -- +(-45:0.4);
  \end{tikzpicture}
}

\newcommand{\DBoxAfterB}{
  \begin{tikzpicture}
    [baseline=2.5ex, x=6ex, y=6ex]
    \coordinate (v1) at (1,0);
    \coordinate (v2) at (1,1);
    \coordinate (v3) at (0,0.5);

    \draw[massless] (v1) -- (v2) -- (v3) -- cycle;

    \draw[massless] (v1) -- +(-45:0.4);
    \draw[massless] (v2) -- +(45:0.4);
    \draw[massless] (v3) -- +(135:0.4);
    \draw[massless] (v3) -- +(-135:0.4);
  \end{tikzpicture}
}

\newcommand{\DBoxBeforeC}{
  \begin{tikzpicture}
    [baseline=-3.5ex, x=6ex, y=6ex]
    \coordinate (v1) at (0,0);
    \coordinate (v2) at (1,0);
    \coordinate (v3) at (2,0);
    \coordinate (v4) at (0,-1);
    \coordinate (v5) at (1,-1);
    \coordinate (v6) at (2,-1);

    \draw[massless] (v5) -- (v2);
    \draw[reduced] (v5) -- (v4) -- (v1) -- (v2) -- (v3) -- (v6) -- cycle;

    \draw[massless] (v4) -- +(-135:0.4);
    \draw[massless] (v1) -- +(135:0.4);
    \draw[massless] (v3) -- +(45:0.4);
    \draw[massless] (v6) -- +(-45:0.4);
  \end{tikzpicture}
}

\newcommand{\DBoxAfterC}{
  \begin{tikzpicture}
    [baseline=-0.5ex, x=6ex, y=6ex]
    \draw[massless] (0,0) circle[radius=0.5];
    \draw[massless] (-0.5,0) -- +(165:0.4);
    \draw[massless] (-0.5,0) -- +(135:0.4);
    \draw[massless] (-0.5,0) -- +(-165:0.4);
    \draw[massless] (-0.5,0) -- +(-135:0.4);
  \end{tikzpicture}
}

\newcommand{\NplDBox}{
  \begin{tikzpicture}
    [baseline=0ex, x=6ex, y=6ex]
    \coordinate (v1) at (1.7,0);
    \coordinate (v2) at (0,0);
    \coordinate (v3) at (0,-1);
    \coordinate (v4) at (1.7,-1);
    \coordinate (v5) at (2.2,-0.5);
    \coordinate (v6) at (1.2,-0.5);

    \draw[massless] (v1) -- (v2) -- (v3) -- (v4);
    \draw[massless] (v1) -- (v5) -- (v4) -- (v6) -- cycle;
    \draw[massless, loop arrow] (v4) -- (v3) node[midway, below]{\(\ell_1\)};
    \draw[massless, loop arrow] (v5) -- (v4);
    \draw (2.2,-1) node{\(\ell_2\)};

    \draw[massless] (v3) -- +(-135:0.4) node[left]{\(k_1\)};
    \draw[massless] (v2) -- +(135:0.4) node[left]{\(k_2\)};
    \draw[massless] (v6) -- +(180:0.4) node[left]{\(k_3\)};
    \draw[massless] (v5) -- +(0:0.4) node[right]{\(k_4\)};
  \end{tikzpicture}
}

\newcommand{\NplDBoxBeforeA}{
  \begin{tikzpicture}
    [baseline=-3.5ex, x=6ex, y=6ex]
    \coordinate (v1) at (1.7,0);
    \coordinate (v2) at (0,0);
    \coordinate (v3) at (0,-1);
    \coordinate (v4) at (1.7,-1);
    \coordinate (v5) at (2.2,-0.5);
    \coordinate (v6) at (1.2,-0.5);

    \draw[massless] (v1) -- (v2) -- (v3) -- (v4);
    \draw[reduced] (v1) -- (v5) -- (v4) -- (v6) -- cycle;

    \draw[massless] (v3) -- +(-135:0.4);
    \draw[massless] (v2) -- +(135:0.4);
    \draw[massless] (v5) -- +(0:0.4);
    \draw[massless] (v6) -- +(180:0.4);
  \end{tikzpicture}
}

\newcommand{\NplDBoxBeforeB}{
  \begin{tikzpicture}
    [baseline=-3.5ex, x=6ex, y=6ex]
    \coordinate (v1) at (1.7,0);
    \coordinate (v2) at (0,0);
    \coordinate (v3) at (0,-1);
    \coordinate (v4) at (1.7,-1);
    \coordinate (v5) at (2.2,-0.5);
    \coordinate (v6) at (1.2,-0.5);

    \draw[reduced] (v1) -- (v2) -- (v3) -- (v4);
    \draw[reduced] (v1) -- (v5) -- (v4);
    \draw[massless] (v1) -- (v6) -- (v4);

    \draw[massless] (v3) -- +(-135:0.4);
    \draw[massless] (v2) -- +(135:0.4);
    \draw[massless] (v5) -- +(0:0.4);
    \draw[massless] (v6) -- +(180:0.4);
  \end{tikzpicture}
}

\newcommand{\NplDBoxBeforeC}{
  \begin{tikzpicture}
    [baseline=-3.5ex, x=6ex, y=6ex]
    \coordinate (v1) at (1.7,0);
    \coordinate (v2) at (0,0);
    \coordinate (v3) at (0,-1);
    \coordinate (v4) at (1.7,-1);
    \coordinate (v5) at (2.2,-0.5);
    \coordinate (v6) at (1.2,-0.5);

    \draw[reduced] (v1) -- (v2) -- (v3) -- (v4);
    \draw[massless] (v1) -- (v5) -- (v4);
    \draw[reduced] (v1) -- (v6) -- (v4);

    \draw[massless] (v3) -- +(-135:0.4);
    \draw[massless] (v2) -- +(135:0.4);
    \draw[massless] (v5) -- +(0:0.4);
    \draw[massless] (v6) -- +(180:0.4);
  \end{tikzpicture}
}

\newcommand{\Ladder}{
  \begin{tikzpicture}
    [baseline=-3.5ex, x=6ex, y=6ex]
    \coordinate (t1) at (0,0);
    \coordinate (t2) at (1,0);
    \coordinate (t3) at (2,0);
    \coordinate (t4) at (3.5,0);
    \coordinate (t5) at (4.5,0);
    \coordinate (t6) at (5.5,0);
    \coordinate (b1) at (0,-1);
    \coordinate (b2) at (1,-1);
    \coordinate (b3) at (2,-1);
    \coordinate (b4) at (3.5,-1);
    \coordinate (b5) at (4.5,-1);
    \coordinate (b6) at (5.5,-1);

    \draw[massless] (t1) -- (t2) -- (t3);
    \draw[massless] (t4) -- (t5) -- (t6);
    \draw[massless] (b1) -- (b2) -- (b3);
    \draw[massless] (b4) -- (b5) -- (b6);
    \draw[massless] (t1) -- (b1);
    \draw[massless] (t2) -- (b2);
    \draw[massless] (t3) -- (b3);
    \draw[massless] (t4) -- (b4);
    \draw[massless] (t5) -- (b5);
    \draw[massless] (t6) -- (b6);
    \draw[massless, loop arrow] (b2) -- (b1) node[midway, below]{\(\ell_1\)};
    \draw[massless, loop arrow] (b3) -- (b2) node[midway, below]{\(\ell_2\)};
    \draw[massless, loop arrow] (b5) -- (b4) node[midway, below]{\(\ell_{L-1}\)};
    \draw[massless, loop arrow] (b6) -- (b5) node[midway, below]{\(\ell_L\)};

    \draw[massless] (b3) -- (2.1,-1);
    \draw[dotdot] (2.1,-1) -- (2.5,-1);
    \draw[massless] (t3) -- (2.1,0);
    \draw[dotdot] (2.1,0) -- (2.5,0);
    
    \draw[massless] (3.4,-1) -- (b4);
    \draw[dotdot] (3,-1) -- (3.4,-1);
    \draw[massless] (3.4,0) -- (t4);
    \draw[dotdot] (3,0) -- (3.4,0);

    \draw (2.60,-0.5) node[circle,fill,inner sep=0.65pt]{};
    \draw (2.75,-0.5) node[circle,fill,inner sep=0.65pt]{};
    \draw (2.90,-0.5) node[circle,fill,inner sep=0.65pt]{};

    \draw[massless] (b1) -- +(-135:0.4) node[left]{\(k_1\)};
    \draw[massless] (t1) -- +(135:0.4) node[left]{\(k_2\)};
    \draw[massless] (t6) -- +(45:0.4) node[right]{\(k_3\)};
    \draw[massless] (b6) -- +(-45:0.4) node[right]{\(k_4\)};
  \end{tikzpicture}
}

\newcommand{\Beetle}{
  \begin{tikzpicture}
    [baseline=
    0ex, x=6ex, y=6ex]
    \coordinate (v1) at (0,0);
    \coordinate (v2) at (0,1);
    \coordinate (v3) at (0,2);
    \coordinate (v4) at (2,2);
    \coordinate (v5) at (2,0);
    \coordinate (v6) at (1,0);

    \draw[massless]  (v1) -- (v2) -- (v3) -- (v4) -- (v5) -- (v6) -- cycle;
    \draw[massless] (v2) -- (v6);
    \draw[massless, loop arrow] (v6) -- (v1) node[midway, below]{\(\ell_1\)};
    \draw[massless, loop arrow] (v5) -- (v6) node[midway, below]{\(\ell_2\)};

     \draw[massless] (v1) -- +(-135:0.4) node[left]{\(k_1\)};
     \draw[massless] (v3) -- +(135:0.4) node[left]{\(k_2\)};
     \draw[massless] (v4) -- +(45:0.4) node[right]{\(k_3\)};
     \draw[massless] (v5) -- +(-45:0.4) node[right]{\(k_4\)};
  \end{tikzpicture}
}

\newcommand{\BeetleCollinearPowerlike}{
  \begin{tikzpicture}
    [baseline=
    0ex, x=6ex, y=6ex]
    \coordinate (v1) at (0,0);
    \coordinate (v2) at (0,1);
    \coordinate (v3) at (0,2);
    \coordinate (v4) at (2,2);
    \coordinate (v5) at (2,0);
    \coordinate (v6) at (1,0);

    \draw[massless]  (v3) -- (v4) -- (v5);
    \draw[masslesscollinear] (v2) -- (v6);
    \draw[masslesscollinear] (v1) -- (v2) -- (v3);
    \draw[masslesscollinear] (v1) -- (v6) -- (v5);

     \draw[masslesscollinear] (v1) -- +(-135:0.4) node[left]{\(k_1\)};
     \draw[massless] (v3) -- +(135:0.4) node[left]{\(k_2\)};
     \draw[massless] (v4) -- +(45:0.4) node[right]{\(k_3\)};
     \draw[massless] (v5) -- +(-45:0.4) node[right]{\(k_4\)};
  \end{tikzpicture}
}

\def\doublebox{\tikz[baseline={2*height("$=$")}]{
\draw[line width=0.35ex] (0.5,-0.3) -- (1,0);
\draw[line width=0.35ex] (0.5,1.3) -- (1,1);
\draw[line width=0.35ex] (1,0) -- (1,1);
\draw[line width=0.35ex] (1,0) -- (2,0);
\draw[line width=0.35ex] (2,0) -- (3,0);
\draw[line width=0.35ex] (3,0) -- (3.5,-0.3);
\draw[line width=0.35ex] (2,0) -- (2,1);
\draw[line width=0.35ex] (3,0) -- (3,1);
\draw[line width=0.35ex] (1,1) -- (2,1);
\draw[line width=0.35ex] (2,1) -- (3,1);
\draw[line width=0.35ex] (3,1) -- (3.5,1.3);
}}

\def\doubleboxcrossed{\tikz[baseline=15]{
\draw[line width=0.35ex] (0.5,-0.3) -- (1,0);
\draw[line width=0.35ex] (0.5,2.3) -- (1,2);
\draw[line width=0.35ex] (1,0) -- (1,2);
\draw[line width=0.35ex] (1,0) -- (2,0);
\draw[line width=0.35ex] (2,0) -- (2.5,-0.3);
\draw[line width=0.35ex] (1,1) -- (2,1);
\draw[line width=0.35ex] (2,0) -- (2,2);
\draw[line width=0.35ex] (1,2) -- (2,2);
\draw[line width=0.35ex] (2,2) -- (2.5,2.3);
}}

\def\trianglebox{\tikz[baseline=10]{
\draw[line width=0.35ex] (0.5,0) -- (1,0.5);
\draw[line width=0.35ex] (0.5,1) -- (1,0.5);
\draw[line width=0.35ex] (1,0.5) -- (3,0);
\draw[line width=0.35ex] (3,0) -- (3.5,-0.3);
\draw[line width=0.35ex] (2,0.25) -- (2,0.75);
\draw[line width=0.35ex] (3,0) -- (3,1);
\draw[line width=0.35ex] (1,0.5) -- (3,1);
\draw[line width=0.35ex] (3,1) -- (3.5,1.3);
}}

\def\triangleboxcrossed{\tikz[rotate=90,baseline=47]{
\draw[line width=0.35ex] (0.7,-0.2) -- (1,0.5);
\draw[line width=0.35ex] (0.7,1.2) -- (1,0.5);
\draw[line width=0.35ex] (1,0.5) -- (3,0);
\draw[line width=0.35ex] (3,0) -- (3.4,-0.5);
\draw[line width=0.35ex] (2,0.25) -- (2,0.75);
\draw[line width=0.35ex] (3,0) -- (3,1);
\draw[line width=0.35ex] (1,0.5) -- (3,1);
\draw[line width=0.35ex] (3,1) -- (3.4,1.5);
}}

\def\slashbox{\tikz[baseline={2*height("$=$")}]{
\draw[line width=0.35ex] (0.5,-0.2) -- (1,0);
\draw[line width=0.35ex] (0.5,1.2) -- (1,1);
\draw[line width=0.35ex] (1,0) -- (1,1);
\draw[line width=0.35ex] (1,1) -- (2,1);
\draw[line width=0.35ex] (1,0) -- (2,0);
\draw[line width=0.35ex] (2,0) -- (2.5,-0.2);
\draw[line width=0.35ex] (2,0) -- (2,1);
\draw[line width=0.35ex] (2,1) -- (2.5,1.2);
\draw[line width=0.35ex] (1,0) -- (2,1)
}}

\def\slashboxsix{\tikz[baseline={2*height("$=$")}]{
\draw[line width=0.35ex] (0.5,-0.2) -- (1,0);
\draw[line width=0.35ex] (0.5,1.2) -- (1,1);
\draw[line width=0.35ex] (1,0) -- (1,1);
\draw[line width=0.35ex] (1,1) -- (2,1);
\draw[line width=0.35ex] (1,0) -- (2,0);
\draw[line width=0.35ex] (2,0) -- (2.5,-0.2);
\draw[line width=0.35ex] (2,0) -- (2,1);
\draw[line width=0.35ex] (2,1) -- (2.5,1.2);
\draw[line width=0.35ex] (1,0) -- (1.7,1)
}}

\def\slashboxsixcrossed{\tikz[baseline={2*height("$=$")}]{
\draw[line width=0.35ex] (0.5,-0.2) -- (1,0);
\draw[line width=0.35ex] (0.5,1.2) -- (1,1);
\draw[line width=0.35ex] (1,0) -- (1,1);
\draw[line width=0.35ex] (1,1) -- (2,1);
\draw[line width=0.35ex] (1,0) -- (2,0);
\draw[line width=0.35ex] (2,0) -- (2.5,-0.2);
\draw[line width=0.35ex] (2,0) -- (2,1);
\draw[line width=0.35ex] (2,1) -- (2.5,1.2);
\draw[line width=0.35ex] (2,0) -- (1,0.7)
}}

\def\doubletriangle{\tikz[baseline={2*height("$=$")}]{
\draw[line width=0.35ex] (0.5,-0.3) -- (1,0);
\draw[line width=0.35ex] (0.5,1.3) -- (1,1);
\draw[line width=0.35ex] (1,0) -- (1,1);
\draw[line width=0.35ex] (1,0) -- (2,0.5);
\draw[line width=0.35ex] (2,0.5) -- (3,0);
\draw[line width=0.35ex] (3,0) -- (3.5,-0.3);
\draw[line width=0.35ex] (3,0) -- (3,1);
\draw[line width=0.35ex] (1,1) -- (2,0.5);
\draw[line width=0.35ex] (2,0.5) -- (3,1);
\draw[line width=0.35ex] (3,1) -- (3.5,1.3);
}}

\def\doubletrianglecrossed{\tikz[rotate=90,baseline=50]{
\draw[line width=0.35ex] (0.6,-0.45) -- (1,0);
\draw[line width=0.35ex] (0.6,1.45) -- (1,1);
\draw[line width=0.35ex] (1,0) -- (1,1);
\draw[line width=0.35ex] (1,0) -- (2,0.5);
\draw[line width=0.35ex] (2,0.5) -- (3,0);
\draw[line width=0.35ex] (3,0) -- (3.4,-0.45);
\draw[line width=0.35ex] (3,0) -- (3,1);
\draw[line width=0.35ex] (1,1) -- (2,0.5);
\draw[line width=0.35ex] (2,0.5) -- (3,1);
\draw[line width=0.35ex] (3,1) -- (3.4,1.45);
}}

\def\teepee{\tikz[baseline={2*height("$=$")}]{
\draw[line width=0.35ex] (0.5,-0.3) -- (1,0);
\draw[line width=0.35ex] (3,0) -- (3.5,-0.3);
\draw[line width=0.35ex] (1,0) -- (3,0);
\draw[line width=0.35ex] (2,0) -- (2,1);
\draw[line width=0.35ex] (2,1) -- (1,0);
\draw[line width=0.35ex] (2,1) -- (3,0);
\draw[line width=0.35ex] (2,1) -- (1.3,1.2);
\draw[line width=0.35ex] (2,1) -- (2.7,1.2);
}}

\def\teepeecrossed{\tikz[rotate=90,baseline=50]{
\draw[line width=0.35ex] (0.55,-0.65) -- (1,0);
\draw[line width=0.35ex] (3,0) -- (3.45,-0.65);
\draw[line width=0.35ex] (1,0) -- (3,0);
\draw[line width=0.35ex] (2,0) -- (2,1);
\draw[line width=0.35ex] (2,1) -- (1,0);
\draw[line width=0.35ex] (2,1) -- (3,0);
\draw[line width=0.35ex] (2,1) -- (1.4,1.5);
\draw[line width=0.35ex] (2,1) -- (2.6,1.5);
}}

\newcommand{\SBoxRegionA}{
  \begin{tikzpicture}
    [baseline=-3.5ex, x=6ex, y=6ex]
    \coordinate (v1) at (0,0);
    \coordinate (v2) at (1,0);
    \coordinate (v3) at (1,-1);
    \coordinate (v4) at (0,-1);

    \draw[bigmassless,teal, dotted] (v1) -- (v2);
    \draw[bigmassless] (v3) -- (v4);
    \draw[bigmassless]  (v2) -- (v3);
    \draw[bigmassless,teal, dotted] (v4) -- (v1);

    \draw[bigmassless] (v4) -- +(-135:0.4) ;
    \draw[bigmassless,teal, dotted] (v1) -- +(135:0.4) ;
    \draw[bigmassless] (v2) -- +(45:0.4) ;
    \draw[bigmassless] (v3) -- +(-45:0.4) ;
  \end{tikzpicture}
}

\newcommand{\SBoxRegionB}{
  \begin{tikzpicture}
    [baseline=-3.5ex, x=6ex, y=6ex]
    \coordinate (v1) at (0,0);
    \coordinate (v2) at (1,0);
    \coordinate (v3) at (1,-1);
    \coordinate (v4) at (0,-1);

    \draw[bigmassless,teal, dotted] (v1) -- (v2);
    \draw[bigmassless] (v3) -- (v4);
    \draw[bigmassless,teal, dotted]  (v2) -- (v3);
    \draw[bigmassless] (v4) -- (v1);

    \draw[bigmassless] (v4) -- +(-135:0.4) ;
    \draw[bigmassless] (v1) -- +(135:0.4) ;
    \draw[bigmassless,teal, dotted] (v2) -- +(45:0.4) ;
    \draw[bigmassless] (v3) -- +(-45:0.4) ;
  \end{tikzpicture}
}

\newcommand{\SBoxRegionC}{
  \begin{tikzpicture}
    [baseline=-3.5ex, x=6ex, y=6ex]
    \coordinate (v1) at (0,0);
    \coordinate (v2) at (1,0);
    \coordinate (v3) at (1,-1);
    \coordinate (v4) at (0,-1);

    \draw[bigmassless] (v1) -- (v2);
    \draw[bigmassless,teal, dotted] (v3) -- (v4);
    \draw[bigmassless,teal, dotted]  (v2) -- (v3);
    \draw[bigmassless] (v4) -- (v1);

    \draw[bigmassless] (v4) -- +(-135:0.4) ;
    \draw[bigmassless] (v1) -- +(135:0.4) ;
    \draw[bigmassless] (v2) -- +(45:0.4) ;
    \draw[bigmassless,teal, dotted] (v3) -- +(-45:0.4) ;
  \end{tikzpicture}
}

\newcommand{\SBoxRegionD}{
  \begin{tikzpicture}
    [baseline=-3.5ex, x=6ex, y=6ex]
    \coordinate (v1) at (0,0);
    \coordinate (v2) at (1,0);
    \coordinate (v3) at (1,-1);
    \coordinate (v4) at (0,-1);

    \draw[bigmassless] (v1) -- (v2);
    \draw[bigmassless,teal, dotted] (v3) -- (v4);
    \draw[bigmassless]  (v2) -- (v3);
    \draw[bigmassless,teal, dotted] (v4) -- (v1);

    \draw[bigmassless,teal, dotted] (v4) -- +(-135:0.4) ;
    \draw[bigmassless] (v1) -- +(135:0.4) ;
    \draw[bigmassless] (v2) -- +(45:0.4) ;
    \draw[bigmassless] (v3) -- +(-45:0.4) ;
  \end{tikzpicture}
}
\newcommand{\SBoxRegionE}{
  \begin{tikzpicture}
    [baseline=-3.5ex, x=6ex, y=6ex]
    \coordinate (v1) at (0,0);
    \coordinate (v2) at (1,0);
    \coordinate (v3) at (1,-1);
    \coordinate (v4) at (0,-1);

    \draw[bigmassless] (v1) -- (v2);
    \draw[bigmassless,red, densely dashed] (v3) -- (v4);
    \draw[bigmassless,teal, dotted]  (v2) -- (v3);
    \draw[bigmassless,teal, dotted] (v4) -- (v1);

    \draw[bigmassless,teal, dotted] (v4) -- +(-135:0.4) ;
    \draw[bigmassless] (v1) -- +(135:0.4) ;
    \draw[bigmassless] (v2) -- +(45:0.4) ;
    \draw[bigmassless,teal, dotted] (v3) -- +(-45:0.4) ;
  \end{tikzpicture}
}

\newcommand{\SBoxRegionF}{
  \begin{tikzpicture}
    [baseline=-3.5ex, x=6ex, y=6ex]
    \coordinate (v1) at (0,0);
    \coordinate (v2) at (1,0);
    \coordinate (v3) at (1,-1);
    \coordinate (v4) at (0,-1);

    \draw[bigmassless,teal, dotted] (v1) -- (v2);
    \draw[bigmassless,teal, dotted] (v3) -- (v4);
    \draw[bigmassless,red, densely dashed]  (v2) -- (v3);
    \draw[bigmassless] (v4) -- (v1);

    \draw[bigmassless] (v4) -- +(-135:0.4) ;
    \draw[bigmassless] (v1) -- +(135:0.4) ;
    \draw[bigmassless,teal, dotted] (v2) -- +(45:0.4) ;
    \draw[bigmassless,teal, dotted] (v3) -- +(-45:0.4) ;
  \end{tikzpicture}
}

\newcommand{\SBoxRegionG}{
  \begin{tikzpicture}
    [baseline=-3.5ex, x=6ex, y=6ex]
    \coordinate (v1) at (0,0);
    \coordinate (v2) at (1,0);
    \coordinate (v3) at (1,-1);
    \coordinate (v4) at (0,-1);

    \draw[bigmassless,red, densely dashed] (v1) -- (v2);
    \draw[bigmassless] (v3) -- (v4);
    \draw[bigmassless,teal, dotted]  (v2) -- (v3);
    \draw[bigmassless,teal, dotted] (v4) -- (v1);

    \draw[bigmassless] (v4) -- +(-135:0.4) ;
    \draw[bigmassless,teal, dotted] (v1) -- +(135:0.4) ;
    \draw[bigmassless,teal, dotted] (v2) -- +(45:0.4) ;
    \draw[bigmassless] (v3) -- +(-45:0.4) ;
  \end{tikzpicture}
}

\newcommand{\SBoxRegionH}{
  \begin{tikzpicture}
    [baseline=-3.5ex, x=6ex, y=6ex]
    \coordinate (v1) at (0,0);
    \coordinate (v2) at (1,0);
    \coordinate (v3) at (1,-1);
    \coordinate (v4) at (0,-1);

    \draw[bigmassless,teal, dotted] (v1) -- (v2);
    \draw[bigmassless,teal, dotted] (v3) -- (v4);
    \draw[bigmassless]  (v2) -- (v3);
    \draw[bigmassless,red, densely dashed] (v4) -- (v1);

    \draw[bigmassless,teal, dotted] (v4) -- +(-135:0.4) ;
    \draw[bigmassless,teal, dotted] (v1) -- +(135:0.4) ;
    \draw[bigmassless] (v2) -- +(45:0.4) ;
    \draw[bigmassless] (v3) -- +(-45:0.4) ;
  \end{tikzpicture}
}

\usepackage{amsmath}
\usepackage{mathtools}
\usepackage{bm} 

\usepackage{array, booktabs, makecell, multirow}
\usepackage[colorlinks=true]{hyperref}
\hypersetup{
  linkcolor=NavyBlue,
  citecolor=NavyBlue,
  filecolor=Red,
  urlcolor=NavyBlue,
  menucolor=Red,
  runcolor=Red
}

\usepackage{cleveref} 

\newcommand{\eps}{\epsilon}
\newcommand{\Ord}{\mathcal{O}}

\renewcommand*{\arraystretch}{.7}
\def\Gram#1#2{G\begin{pmatrix} #1\\ #2 \end{pmatrix}}
\def\GramO#1{G\begin{pmatrix} #1\end{pmatrix}}
\def\LLort#1#2{\nu_{#1#2}}
\def\Lort#1{\hat{\ell}_{#1}}
\def\FlowSymb{\beta}
\def\FlowSymbX{\gamma}
\allowdisplaybreaks

\begin{document}

\title{Finite Feynman Integrals}

\author{Giulio Gambuti}
\affiliation{Rudolf Peierls Centre for Theoretical Physics, University of Oxford, Clarendon Laboratory, Parks Road, Oxford OX1 3PU, UK\\
\textsf{\rm\sf Giulio.Gambuti@new.ox.ac.uk}}

\author{David A. Kosower}
\affiliation{Institut de Physique Théorique, CEA, CNRS, Université Paris--Saclay, F--91191 Gif-sur-Yvette cedex, France\\
\textrm{\rm and} Theoretical Physics Department, CERN, Geneva, Switzerland\\
\textsf{\rm\sf David.Kosower@ipht.fr}}

\author{Pavel P. Novichkov}
\affiliation{Institut de Physique Théorique, CEA, CNRS, Université Paris--Saclay, F--91191 Gif-sur-Yvette cedex, France\\
\textsf{\rm\sf Pavel.Novichkov@ipht.fr}}

\author{Lorenzo Tancredi}
\affiliation{Physik Department, James-Franck-Straße 1, Technische Universität München, D--85748 Garching, Germany\\
\textsf{\rm\sf Lorenzo.Tancredi@tum.de}}

\date{\today}

\preprint{TUM-HEP-1483/23}
\preprint{OUTP-23-13P}
\preprint{CERN-TH-2023-218}

\begin{abstract}
We describe an algorithm to organize Feynman integrals
in terms of their infra-red properties.
Our approach builds upon the theory of Landau singularities,
which we use to classify all configurations of loop momenta that can give rise
to infrared divergences. We then construct
bases of numerators for arbitrary Feynman integrals, which cancel
all singularities and render the integrals finite. Through the same analysis,
one can also classify so-called evanescent and evanescently finite
Feynman integrals.  These are integrals whose vanishing or finiteness
relies on properties of dimensional regularization.
To illustrate the use of these integrals,
we display how to obtain a simpler form for
the leading-color two-loop four-gluon scattering amplitude through
the choice of a suitable basis of finite integrals.
In particular, when all gluon helicities are equal, we show that with our basis
the most complicated double-box integrals do not contribute to the finite remainder
of the scattering amplitude.

\end{abstract}

\maketitle

\tableofcontents

\section{Introduction}

Feynman integrals are key building blocks for observables in 
high-energy physics.  The integrals contributing to any scattering amplitude
satisfy many linear relations.
Within dimensional regularization~\cite{Bollini:1972ui,tHooft:1972tcz}, 
integration-by-parts identities (IBPs) \cite{Tkachov:1981wb,Chetyrkin:1981qh} 
provide these relations systematically.  Solving these relations
allows us to express scattering amplitudes in terms
of a basis of so-called master integrals.  As with the basis in
any linear space, the choice of master integrals is not unique.
Different choices 
may be appropriate for different purposes. For instance,  canonical 
bases \cite{Arkani-Hamed:2010pyv,Kotikov:2012ac,Henn:2013pwa} are 
particularly well-suited to the computation of the integrals themselves via 
the method of differential equations 
\cite{Kotikov:1990kg,Bern:1993kr,Remiddi:1997ny,Gehrmann:1999as}.
However such bases are not necessarily the best option when seeking 
a compact representation for scattering amplitudes.  The choice
of an optimal basis for general
multiloop scattering amplitudes is not known; indeed, even the
criteria for such a 
representation are a matter of debate.

In this article, we take the first step towards answering this question,
with Yang--Mills and gravity theories in mind.
These theories exhibit both infrared (IR) and ultraviolet (UV) divergences.
In dimensional regularization both appear as poles in the regulator 
$\epsilon$. The structure of IR divergences in scattering amplitudes is 
predicted by IR factorization~\cite{Catani:1998bh, Becher:2009cu,Becher:2009qa,Almelid:2015jia,Agarwal:2021ais}. 
In practice, an $L$-loop amplitude 
can be decomposed into hard and IR-divergent parts,
$\mathcal{A}^{(L)} =\mathcal{A}_\mathrm{hard}^{(L)} + 
\mathcal{A}_\text{IR}^{(L)}$, 
where  $\mathcal{A}_\text{IR}$ contains all IR divergences of the 
amplitude.  The latter terms can be universally predicted from 
lower-loop information, 
whereas \( \mathcal{A}_\mathrm{hard}^{(L)} \) captures process-dependent 
intrinsically $L$-loop information. 

The IR factorization of scattering amplitudes offers a natural path to 
organize bases of Feynman integrals. As an example, 
the authors of refs.~\cite{Henn:2019rmi} 
and~\cite{vonManteuffel:2020vjv}, following two different approaches, 
simplified the calculation of the four-loop QCD cusp anomalous dimension. 
Each chose a basis of master integrals that minimized the number and complexity 
of IR-divergent integrals required. 
IR-finite integrals have also been employed~\cite{Bonetti:2020hqh,Bonetti:2022lrk}
to simplify the calculation of the 
mixed QCD-EW corrections to the production 
of a Higgs boson and a jet. 
These finidings support the idea that organizing integrals according to their IR properties has 
the potential to simplify not only the final form of the result 
but also the computation of scattering amplitudes. 
The most straightforward way of making the IR structure of the amplitude manifest is
having the hard scattering part described by IR-finite integrals. This sets
our immediate goal: we focus on the classification of finite Feynman
integrals as candidates for representing $\mathcal{A}_\text{hard}^{(L)}$. 
These integrals fall into three different classes:
\textit{locally finite}, \textit{evanescent} and \textit{evanescently finite}. 

Locally finite integrals have integrands whose singularities are integrable. 
They can be computed directly in four dimensions and are in that sense more
robustly finite than integrals in the third category. 
Von Manteuffel, Panzer, and Schabinger previously
developed~\cite{vonManteuffel:2014qoa,vonManteuffel:2015gxa}
an approach to constructing bases of finite integrals. Their
approach relied on the study of Feynman integrals in parametric representation, 
where their finiteness can be determined from the degree of divergence in the 
individual integration parameters.
Finite integrals can then be determined by adding powers of the individual 
propagators and considering them in different numbers of space-time dimensions.
While this approach is strictly guarateed to work only in the Euclidean case, 
it has already found application in various state-of-the-art 
calculations~\cite{Borowka:2016ehy,Borowka:2016ypz,vonManteuffel:2017myy,%
Chen:2020gae}. 
Finite integrals are simpler to evaluate.  It is easier to compute them
analytically by direct integration~\cite{Panzer:2014caa}. They also are
easier to 
evaluate to high precision using purely numerical methods
either with sector decomposition~\cite{vonManteuffel:2017myy} or
methods based on tropical geometry~\cite{Borinsky:2023jdv}.

Evanescent integrals are a subset of locally finite integrals which vanish in four
dimensions, but are non-trivial in $D$ dimensions.  That is, they evaluate to
$\Ord(\eps)$ and therefore do not contribute to the finite part of scattering 
amplitudes at the corresponding loop order (so long as there are no factors of 
$1/\eps$ in their coefficients).  Instead, they give rise to new linear relations
between basis integrals up to corrections which vanish in a physical observable.  
These integrals have already found application within the method of differential
equations~\cite{Remiddi:2013joa,Duhr:2022dxb}, where
they can be used to decouple systems near even integer dimensions.

Evanescently finite integrals feature a cancellation between an otherwise
evanescent numerator (vanishing in four dimensions) and either UV or IR divergences,
so that the singularities of the integrand are not locally integrable, but after 
integration one still obtains a finite result. Both evanescent and evanescently 
finite integrals are special to dimensional regularization, as both classes of 
integrals are ill-defined in four dimensions.  In contrast, locally finite 
integrals are well-defined in four dimensions, and are expected to be independent 
of the UV and IR regulators.
In order to isolate finite integrals, we make use of Weinberg's 
theorem~\cite{Dyson:1949ha,Weinberg:1959nj} to establish UV finiteness and on the 
time-honored approach of the Landau equations~\cite{Landau:1959fi,*Bjorken:1959fd,*Nakanishi:1959} to 
systematically identify all possible sources of IR divergences%
\footnote{The topic has recently enjoyed renewed interest, with
exploration of new approaches to determining the locus of singular
points associated to a given Feynman
diagram~\cite{Mizera:2021icv,Hannesdottir:2021kpd,Gardi:2022khw}.}.


The rest of this paper is organized as follows. In \cref{sec:nota}
we review notation and ingredients for later sections.
In \cref{sec:weinberg_uv}, we review Weinberg's theorem and conditions for
UV finiteness.  In \cref{sec:IR} we present well-known aspects of the theory of Landau
singularities and then present our general algorithm to solve them and identify
finite, evanescent and evanescently finite integrals.
In \cref{sec:onebox} we explain our approach using the one-loop box
as an example.
We continue in \cref{sec:evanescent} with a discussion of evanescent and 
evanescently finite integrands in more detail.  We use the one-loop pentagon
as an example.
Our main results are collected in \cref{sec:results}, where we show how to 
apply our
techniques to increasingly complicated cases, up to four loops. In particular,
we also demonstrate how choosing a basis of finite integrals helps make
the singularity structure of the leading-color all-plus four-gluon amplitude
manifest. We summarize in \cref{sec:conclusions}.
Explicit formulas can be found in the appendices.

\section{Notation}%
\label{sec:nota}

We write an $L$-loop Feynman integral in dimensional regularization as,
\begin{equation}
  \label{eq:feynman_integral}
I[\mathcal{N}(\ell_i)] =
  \int \prod_{i=1}^{L} \mathrm{d}^D \ell_i \, \frac{\mathcal{N}(\ell_i)}{\mathcal{D}_1 \cdots \mathcal{D}_E},
\end{equation}
where $\mathcal{D}_e = q_e^2 - m_e^2 + i \varepsilon$ are the $E$
propagators of the 
corresponding graph (with $E$ edges)
and $\mathcal{N}$ is a Lorentz-invariant numerator.
The edge momenta~\(q_e\) are linear combinations of the loop momenta~\(\ell_i\) 
and the external momenta~\(k_j\) (\(j = 1, \dots, n\)) with 
coefficients \(\pm 1\) or 0.  We consider integrals near four dimensions,
$D=4-2\eps$.
The problem of classifying all finite integrals for a given graph (also called 
\emph{topology}) can then be solved by finding all numerators $\mathcal{N}$ that 
make the integral~\eqref{eq:feynman_integral} finite.
Our goal is to identify the general form of $\mathcal{N}$.

Lorentz invariance requires numerators of Feynman integrals to be built out of 
scalar products of all available vectors: the loop momenta $\ell_i$, the external 
momenta $k_j$, and any other external vectors $Q_j$ independent 
of the loop momenta
appearing in scattering amplitudes, such as polarization vectors. We take all
external vectors to be strictly four-dimensional. 
Numerators of Feynman integrals are then polynomials in $\ell_i \cdot \ell_j$, 
$\ell_i \cdot k_j$ and $\ell_i \cdot Q_j$ with coefficients 
that are rational functions of
the external kinematic invariants. 
We can reduce scalar products of the form
$\ell_i \cdot Q_j$ to combinations of $\ell_i \cdot \ell_j$ and $\ell_i \cdot k_j$
through a combination of standard tensor reduction and use of a basis
for external vectors (see ref.~\cite{Anastasiou:2023koq} for recent improvements 
based on the use of the van~Neerven--Vermaseren basis, and 
refs.~\cite{Peraro:2019cjj,Peraro:2020sfm} for a simplified approach that exploits 
four-dimensional external states).  This allows us to write any numerator in
the form,
\begin{equation}
  \label{eq:general_ansatz}
  \mathcal{N}(\ell_i) = \sum_{\vec{r}} c_{\vec{r}} \prod_{a} t_a^{r_a}\,,
\end{equation}
where the $t_a$ belong to the list of allowed scalar products described above,
\begin{equation}
\label{eq:monomials}
    t_a \in  \{ \ell_i \cdot k_j\} \cup \{ \ell_i \cdot \ell_j\}\,,
\end{equation}
and $\vec{r}$ are vectors of non-negative integer numbers representing the powers 
of each monomial. The coefficients $c$ are rational functions of the
kinematic invariants. 
In this article, we use the notation
\(k_{i \dots j} \equiv k_i + \dots + k_j\),
and use as kinematic invariants the Mandelstam variables 
$s_{i \dots j} = k_{i\dots j}^2$ alongside the distinct non-zero masses: external~$m_i^2 = k_i^2$ and internal~\(m_e^2\).
Some masses may vanish, or be equal to other masses, as is the case for integrals 
arising in amplitudes.\footnote{In the literature on Feynman integrals, such 
configurations are sometimes called \enquote{exceptional/non-generic kinematics}.}
In a slight abuse of language, we will refer to the total degree in the loop
momenta as the rank of a term (rather than to the number of free indices);
the rank of a numerator expression will be the maximum rank of any term.

The numerator representation given by \cref{eq:general_ansatz,eq:monomials} 
is not unique. For our purposes, it will be useful to consider an alternative 
representation built using a van Neerven--Vermaseren basis, which we define below. 
We first define the generalized Gram determinant of two sets of $R$ vectors in $D$ 
dimensions,
\begin{equation}
\begin{aligned}
   \Gram{p_1  & \cdots & p_R}{q_1 & \cdots & q_R} &\equiv 
     \mathrm{det}(2p_i\cdot q_j)\,,\\ 
   \GramO{p_1  & \cdots & p_R}&\equiv 
     \Gram{p_1  & \cdots & p_R}{p_1 & \dots & p_R}\,.
   \end{aligned}
   \label{eq:gram_def}
\end{equation}
We define a Gram with a free index,
\begin{equation}
    \Gram{p_1  & \cdots\mu\cdots & p_R}{q_1 & \dots & q_R} \equiv
    \frac{\partial}{\partial w_\mu}
        \Gram{p_1  & \cdots w\cdots & p_R}{q_1 & \dots & q_R} \,.
\end{equation}
Next, suppose that the space of external momenta for an \(n\)-point process is spanned by a basis \(k_1, \dots, k_R\), where \(R = \min (n-1, 4)\) in four dimensions.
The well-known van Neerven--Vermaseren basis can be defined as,
\begin{equation}
  v_i^{\mu} \equiv \frac{\Gram{k_1  & \dots&\mu&\dots & k_R}
   {k_1 & \dots&k_i&\dots & k_R}}{\GramO{k_1 & \dots & k_R}} \,,
   \label{eq:vannver}
\end{equation}
which has the important property,
\begin{equation}
    v_i \cdot k_j = \delta_{ij} \quad \mbox{for}\; 1 \leq j \leq R\,,
\end{equation}
and allows us to decompose any external momentum $w^\mu$ as,
\def\wOrthog{\hat w}
\begin{equation}\label{eq:vNV_decomposition}
    w^{\mu} = \sum_{j=1}^R (v_j \cdot w) \: k_j^\mu + \wOrthog^{\mu} \, ,
\end{equation}
with $k_i \cdot \wOrthog = 0$ for $i=1,\dots,R$. The remainder $\wOrthog$ satisfies $\wOrthog^2 \leq 0$. In the rest of this manuscript we will reserve the ``hat" symbol for out-of-plane components of momenta. 

We also introduce $\Lort{i}$, the out-of-plane part of the loop momentum $\ell_i$ orthogonal
to all external momenta $k_j$.  When considering an $n$-point process with
$n>4$, this part is strictly $\eps$-dimensional.
Using \cref{eq:vNV_decomposition} we can then write the scalar products in \cref{eq:monomials} as linear combinations of those in the set,
\begin{equation}
\label{eq:monomials_alt}
    t_a \in \{ \ell_i \cdot v_j\} \cup  \{ \Lort{i} \cdot \Lort{j} \} \, ,
\end{equation}
which span the same space as the one generated by \cref{eq:monomials} but greatly 
simplify the identification of IR-finite numerators.  This alternate basis
will allow us to generalize our analysis more easily
than the standard basis in \cref{eq:monomials}. 
It is convenient to define the notation,
\begin{equation}
  \LLort{i}{j} \equiv \Lort{i} \cdot \Lort{j} = 
  \Gram{l_i & k_1 & \dots & k_R}{l_j & k_1 & \dots & k_R}\,.
\end{equation}
Conveniently, all definitions above can be continued to $D$ continuous space-time 
dimensions, whenever necessary.

With the notation and numerator representations in hand, we turn to the
constraints on \cref{eq:general_ansatz} imposed by UV and IR finiteness.
We consider UV constraints in the next section, and IR constraints in
following sections.

\section{UV Divergences and Weinberg's Theorem}%
\label{sec:weinberg_uv}
The UV divergences of a Feynman integral arise from regions where
loop momenta become large.
The convergence criterion known today as
Weinberg's theorem was introduced by Dyson~\cite{Dyson:1949ha} and proven 
rigorously by Weinberg~\cite{Weinberg:1959nj} in the case of Euclidean metric.
A simpler proof and an extension to the Minkowski case was presented 
by Hahn and Zimmermann~\cite{Hahn:1968,*Zimmermann:1968}, while 
Nakanishi gave a proof based on the parametric 
representation~\cite{Nakanishi:1957,*Nakanishi:1971}.

The power-counting theorem states that a Feynman integral is UV-finite if 
the integral itself as well as all its subintegrations have a negative 
superficial degree of divergence. 
Subintegration here means integrating over at most \((L \!-\! 1)\)
independent loops of the graph while keeping the remaining propagator momenta 
fixed to generic values.

For any given subintegration, we can always redefine the loop-momentum variables 
so that they
correspond to integration over a proper subset \(S\) of 
the loop momenta~\(\ell_i\) holding the remaining ones fixed. Rescaling 
the unfixed loop momenta in \(S\) as \(\ell^\mu \to \rho \: \ell^\mu \) with
\( \rho \gg 1\), we find 
\begin{equation}
    \mathcal{I} = \Ord(\rho^\omega)\,,
\end{equation}
where \(\omega\) is the superficial degree of divergence of the subintegration.

Below we show explicitly how Weinberg's theorem can be used in practice to 
constrain a generic integral like the one in \cref{eq:feynman_integral}. 
\begin{figure}[tb]
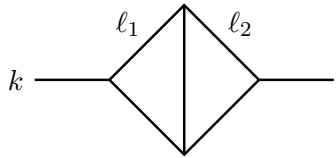

    \centering
    \UVdoubleTriangle
    \caption{\label{fig:example_UV}%
      A simple two-loop two-point integral.}
\end{figure}
We take the simple 2-loop \enquote{kite} integral shown in
\cref{fig:example_UV} as an example. 

Requiring the superficial convergence (\(\omega<0\)) of the whole integral 
puts an upper bound on the numerator rank:
\begin{equation}
  \label{eq:UV_rank_bound}
  \operatorname{rank} \left( \mathcal{N}\right) \leq r_{\mathrm{max}} =
  2 E - 4 L - 1\,.
\end{equation}
For \cref{fig:example_UV}, the rescaling \( \ell_{1,2} \to \rho \; 
\ell_{1,2}\) gives a superficial degree of divergence 
$\omega =  \operatorname{rank} \left( \mathcal{N}\right) - 2$, which implies 
$ \operatorname{rank} \left( \mathcal{N}\right) \leq 1 $:
the numerator can be at most linear in the loop momenta.

Requiring convergence of subintegrations puts additional constraints on 
the numerator.
To find them, we first need to identify all possible subintegrations. 
For simple examples like the one considered here, it is convenient to work in momentum space
and what follows is one way of doing this algorithmically.\footnote{As we shall 
 discuss in \cref{sec:landau_solve}, we analyze more complicated integrals 
in Feynman parameter space. One can then solve the equations obtained by setting the 
 first Symanzyk polynomial to zero monomial by monomial in order to identify the 
 \enquote{unfixed} edges. }


\begin{figure}[tb]
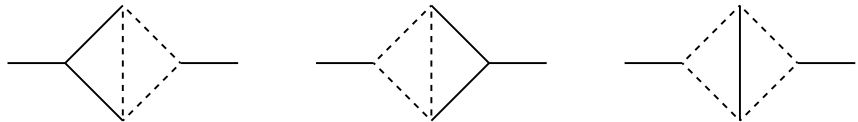

  \centering
  \resizebox{0.2\textwidth}{!}{\UVdoubleTriangleSubA}\, \quad \;
  \resizebox{0.2\textwidth}{!}{\UVdoubleTriangleSubB}\, \quad \;
  \resizebox{0.2\textwidth}{!}{\UVdoubleTriangleSubC}\,
  \caption{\label{fig:example_UV-subintegrations}%
    Graphical depiction of subintegrations for the integral in
    \cref{fig:example_UV}. The active subintegrations are indicated by
    dashed lines, while the momenta corresponding to solid lines are 
    kept fixed.}
\end{figure}
\newcommand{\Lsubset}{V}
We start by listing all distinct linear combinations 
$c^{(v)}_i\ell_i$ of loop momenta entering the denominators (equivalently,
the set of all propagator momenta evaluated at vanishing internal masses 
and external momenta). 
We then pick a subset $\Lsubset$ of these linear combinations 
\(\{ c^{(v)}_i\ell_i\in \Lsubset\}\),
and fix each element in the subset to a different constant,
\begin{equation}
\label{eq:constraints_sub_int}
    c^{(v)}_i\ell_i = d_v\, , \quad \forall v \in \Lsubset \, ,
\end{equation}
and solve the corresponding  system of equations for the loop momenta.
We retain only subsets which leave at least one \(\ell_i\) unconstrained.
In the kite integral, the different linear combinations of loop momenta are 
\(\{\ell_1,\ell_2,\ell_1+\ell_2\}\), while the list of subsets and their corresponing constraints is,
\begin{equation}
\begin{split}
   &\{\ell_1=d_1,\;\ell_2=d_2\} \; , \quad
   \{\ell_1=d_1,\;\ell_1+\ell_2=d_2\} \; , \quad
   \{\ell_2=d_1,\;\ell_1+\ell_2=d_2\} \; ,\\
  & \quad\quad\quad\quad\quad\quad\quad 
  \{\ell_1=d\} \; , \quad
  \{\ell_2=d\} \; , \quad
  \{\ell_1+\ell_2=d\} \; .
   \end{split}
\end{equation}
The subsets in the top row leave no loop momentum unconstrained. Conversely, 
each of the subsets in the bottom row leaves one degree of freedom unfixed and 
therefore corresponds to a subintegration. 
Graphically, these three subsets can be associated to the diagrams
shown in \cref{fig:example_UV-subintegrations}.

With the set of subintegrations at hand, we turn to finding the 
corresponding additional constraints on the numerator.
We write down the most general ansatz with the maximal allowed overall rank
$r_{\mathrm{max}}$ as a linear combination of all possible monomials.
This is equivalent to setting to zero all coefficients $c_{\Vec{r}}$ with 
$\sum_a r_a > r_{\mathrm{max}}$ in the general ansatz of
\cref{eq:general_ansatz}.  It turns the infinite-dimensional space 
of polynomials into a finite-dimensional space of superficially 
UV-convergent numerators:
\begin{equation}
  \label{eq:UV_ansatz}
   \sum_{\vec{r} } c_{\vec{r}} \prod_{a} t_a^{r_a}\qquad
   \text{where\ } \sum_a r_a \leq r_\mathrm{max}\, .
\end{equation}
For each subintegration we substitute the corresponding
solution in \cref{eq:constraints_sub_int} into the general numerator, and 
rescale all unfixed loop momenta as $\ell \to \rho \: \ell$. 
We then expand the numerator in the scaling parameter~\(\rho\), and 
require that the divergent orders in~\(\rho\) vanish.
This yields a system of linear equations.  Solving them for 
\(c_{\vec{r}}\) gives the most general numerator consistent 
with UV finiteness.
In the kite example, the set of monomials is 
\(\{\ell_1 \cdot \ell_2 , \ell_1 \cdot k, \ell_2 \cdot k\}\),
which give the superficially UV-convergent numerator,
\begin{equation}
\label{eq:example_UV_num}
    \mathcal{N} = c_{(0,1,0)} \: \ell_1 \cdot k \; + \;
    c_{(0,0,1)} \: \ell_2 \cdot k  \; .
\end{equation}
The three subintegrations of \cref{fig:example_UV-subintegrations} 
provide respectively the scaling rules,
\begin{equation}
      \{\ell_1=d ,\; \ell_2 \to \rho \:\ell_2\} \; , \quad
  \{ \ell_1 \to \rho \:\ell_1,\;\ell_2=d\} \; , \quad
  \{\ell_1=d-\ell_2,\; \ell_2 \to \rho \:\ell_2\} \;,
\end{equation}
which impose no further constraint on the numerator
in \cref{eq:example_UV_num}.


\section{IR Divergences, Landau Equations and Power Counting}%
\label{sec:IR}
\subsection{Potential Singularities: the Landau Equations}

Infrared divergences of Feynman integrals are associated to 
loop-momentum configurations in which a subset of the propagators
(equivalently a subset of the denominators) in \cref{eq:feynman_integral}, 
vanish. In general, the domain of integration of an $L$-loop Feynman 
integral~\eqref{eq:feynman_integral} with $E$ propagators will contain 
IR-divergent surfaces of various dimensions. These correspond to solutions of
the Landau equations~\cite{Landau:1959fi,*Bjorken:1959fd,*Nakanishi:1959} 
(see also ref.~\cite{Mizera:2021fap} for an up-to-date literature review 
and refs.~\cite{Sterman:1993hfp,*Collins:2011zzd,*Agarwal:2021ais} for a 
pedagogical discussion of IR divergences). 
In presenting the Landau equations, it is convenient to rewrite the integral 
of \cref{eq:feynman_integral} in the mixed representation\footnote{Called the 
\enquote{second representation} in the classic book~\cite{Eden:1966dnq}.}:
\begin{equation}
I[\mathcal{N}(\ell_i)] =
\Gamma(E)  \int \prod_{i=1}^{L} \mathrm{d}^D \ell_i \, 
\int_0^\infty \frac{\prod_{e=1}^E \mathrm{d}\alpha_e}{\mathrm{GL}(1)} \,
\frac{\mathcal{N}(\ell_i) }
{[\alpha_1 \mathcal{D}_1 + \dots + \alpha_E \mathcal{D}_E]^E}\,,
\end{equation}    
where the $\mathrm{GL}(1)$ denominator accounts for the projective invariance 
of the integration over the Feynman parameters $\alpha_e$.  It is equivalent 
to the usual notation $\delta(1 - \sum_{e \in A} \alpha_e)$ with $A$ a 
subset of $\{1,\dots,E\}$. 
Potential singularities\footnote{With the exception of UV singularities which 
we assume have already been removed.} of the integral correspond to 
configurations of the integration variables which satisfy,
\begin{subequations}
  \label{eq:landau}
  \begin{alignat}{2}
    \label{eq:landau_lin}
    \forall i &= 1, \dots, L \colon & \quad 
    \sum_{e=1}^{E} \alpha_e \frac{\partial}{\partial \ell_i} 
     \mathcal{D}_e &= 0\,, \\
    \label{eq:landau_sq}
    \forall e &= 1, \dots, E \colon & \alpha_e \mathcal{D}_e &= 0\,.
  \end{alignat}
\end{subequations}
These are the Landau equations, which should be viewed as a system of 
equations for the loop momenta and the kinematic invariants for some values 
of \(\alpha_e \geq 0\) where at least one \(\alpha_e\) is strictly positive. 

There are two classes of solutions to the Landau equations.
Solutions of the first class impose constraints on the kinematic variables.
Such solutions manifest themselves as singularities of the integrals as a 
function of these variables, and are called \emph{Landau singularities}.
Kinematics-independent solutions, on the other hand, correspond to IR 
divergences, and manifest themselves as poles in the dimensional 
regulator~\(\epsilon\).
In what follows we will only be concerned with the latter class of solutions.

\subsection{Solving the Landau Equations}
\label{sec:landau_solve}

The set of equations~\eqref{eq:landau_lin} are linear in the loop momenta, 
while~\eqref{eq:landau_sq} are quadratic.
We proceed by first solving the linear system~\eqref{eq:landau_lin} 
for the loop momenta, expressing them in terms of the external momenta 
and the \(\alpha\) parameters.  We then substitute the solution into 
the remaining equations~\eqref{eq:landau_sq}, and obtain a system 
of polynomial equations for the \(\alpha\) parameters.  In many cases 
the latter system can be solved easily, as we will see below.

This procedure relates the Landau equations to the Symanzik polynomials of the 
Feynman graph (see refs.~\cite{Nakanishi:1971,Weinzierl:2022eaz}),
\begin{subequations}
  \label{eq:UFdef}
  \begin{align}
    \label{eq:Udef}
    \mathcal{U} (\alpha) &= \sum_{T \in \mathcal{T}_1} 
     \prod_{e \notin T} \alpha_e\,, \\
    \label{eq:Fdef}
    \mathcal{F} (\alpha) &= \sum_{T \in \mathcal{T}_2} k(T)^2
    \prod_{e \notin T} \alpha_e - \mathcal{U} (\alpha) 
    \sum_{e=1}^{E} m_e^2 \alpha_e\,,
  \end{align}
\end{subequations}
where \(\mathcal{T}_n\) is the set of spanning \(n\)-forests of the graph,
and \(k^{\mu}(T)\) is the total momentum flowing across the 2-forest
\(T \in \mathcal{T}_2\).
The Symanzik polynomials can be computed efficiently by rewriting the combined 
denominator in the mixed representation as a quadratic form in the loop 
momenta:
\begin{equation}
  \label{eq:combined_den}
  \sum_{e=1}^{E} \alpha_e \mathcal{D}_e
  = \sum_{i,j=1}^{L} A_{ij} \ell_i \cdot \ell_j
   - 2 \sum_{i=1}^{L} \ell_i \cdot B_i + C\,,
\end{equation}
so that
\begin{subequations}
  \label{eq:UFcomp}
  \begin{align}
    \label{eq:Ucomp}
    \mathcal{U} &= \det A, \\
    \label{eq:Fcomp}
    \mathcal{F} &= \det A \left(-B^T A^{-1} B + C \right).
  \end{align}
\end{subequations}
An immediate consequence of~\cref{eq:combined_den,eq:Ucomp} is that
the linear system~\eqref{eq:landau_lin} is non-degenerate if and only if \(\mathcal{U} \neq 0\).
Let us consider this non-degenerate case first.
As the square matrix~\(A\) is of full rank, the linear 
system~\eqref{eq:landau_lin} admits a unique solution,
\begin{equation}
  \label{eq:landau_lin_sol}
  \ell_i^{\mu} = {\left( A^{-1} \right)}_{ij} B_j^{\mu}\,.
\end{equation}
(A parametrization-independent form of the solution in terms of the graph 
spanning trees can be found in ref.~\cite{Mizera:2021fap}).
We now substitute this solution into the quadratic 
equations~\eqref{eq:landau_sq}, which become simply
\begin{equation}
  \label{eq:landau_par}
  \forall e = 1, \dots, E \colon \quad \alpha_e \frac{\partial}{\partial \alpha_e} \mathcal{F} = 0\,.
\end{equation}
These equations are \emph{also\/} sometimes referred to as \enquote{the Landau equations}; they describe singularities of the integrand 
in the Feynman-parameter representation%
\footnote{\enquote{The third representation} of ref.~\cite{Eden:1966dnq}.}
when \(\mathcal{U} \neq 0\).%
\footnote{A version of \cref{eq:landau_par} which captures also the 
singularities with \(\mathcal{U} = 0\) is obtained by 
replacing~\(\mathcal{F}\) with the 
\emph{worldline action}~\(\mathcal{F} / \mathcal{U}\) suitably continued to the boundary~\cite[Remark~9.2 and Theorem~12.1]{Nakanishi:1971}.}

To solve \cref{eq:landau_par}, observe that, due to Euler's homogeneous
function theorem, \cref{eq:landau_par} implies \(\mathcal{F} = 0\).
In many cases, \(\mathcal{F}\) can be brought into a 
\emph{subtraction-free} form, meaning that each monomial in \(\mathcal{F}\) 
has the same sign in a certain region of the kinematic space
(equivalently, there exists a Euclidean region%
\footnote{The same term is sometimes used to refer to a region where \emph{all} Mandelstam variables 
are positive.  This is not possible in general for scattering processes 
when all external states are massless.}
of the integral).
In particular, this holds for planar Feynman graphs with massless external 
legs.  In these cases the \(k(T)^2\) in~\cref{eq:Fdef} are Mandelstam invariants
of consecutive external momenta (with respect to the planar ordering), 
which can be made negative simultaneously.

For \(\alpha_e \geq 0\), a subtraction-free polynomial \(\mathcal{F}\) can 
only vanish if each monomial in \(\mathcal{F}\) vanishes independently, 
which in turn implies~\cref{eq:landau_par}.
Therefore, for a subtraction-free \(\mathcal{F}\), solving~\cref{eq:landau_par} 
is equivalent to setting all monomials in 
\(\mathcal{F}\) to zero independently.  The solutions have the form,
\begin{equation}
  \label{eq:landau_par_sol}
  \alpha_{e_1} = 0,\;\; \dots,\;\; \alpha_{e_m} = 0\,,
\end{equation}
where \(e_1, \dots, e_m\) is a proper subset of edge labels 
\(\left\{ 1, \dots, E \right\}\).

When \(\mathcal{F}\) does not admit a subtraction-free form, we must solve 
\cref{eq:landau_par} explicitly.
It is believed that the general form of the solution in such cases is
nonetheless also given by~\cref{eq:landau_par_sol}.  That is, IR divergences 
cannot arise due to cancellation between terms in~\(\mathcal{F}\).

One possible strategy to prove this for a given diagram is as follows.
Instead of summing~\cref{eq:landau_par} to get \(\mathcal{F} = 0\), which is not subtraction-free, consider the most general linear combination of~\cref{eq:landau_par}:
\begin{equation}
  \label{eq:landau_par_lin_comb}
  \sum_{e = 1}^E w_e \alpha_e \frac{\partial}{\partial \alpha_e} \mathcal{F} = 0\,.
\end{equation}
The left-hand side of~\cref{eq:landau_par_lin_comb} contains the same monomials as~\(\mathcal{F}\), but the coefficients of these monomials are now linear functions of~\(w_e\).
Demanding that all coefficients are non-zero and have the same sign in some region of the kinematic space, one obtains a system of linear inequalities on~\(w_e\).
If one can find a solution of this system (e.g.\ using numerical linear optimization routines), then solving~\cref{eq:landau_par} is equivalent to setting each monomial in~\(\mathcal{F}\) to zero independently as in the subtraction-free case, therefore no cancellations can occur.
We illustrate this procedure in~\cref{sec:nonplanar_dbox} when we discuss the massless non-planar double box integral.

Having solved the parameter-space Landau equations~\eqref{eq:landau_par}, it remains to check that the solution is consistent with the condition~\(\mathcal{U} \neq 0\), and compute the corresponding loop momenta using~\cref{eq:landau_lin_sol}.

We turn next to the degenerate case, \(\mathcal{U} = 0\).
As we can see from~\cref{eq:Udef}, \(\mathcal{U}\) is always subtraction-free, 
so this condition implies that a subset of \(\alpha_e\) vanishes, and all such 
subsets are identified by setting all monomials in \(\mathcal{U}\) to 
zero.
Each solution of \(\mathcal{U} = 0\) corresponds to the situation where a subdiagram  \(\gamma\) of the original graph~\(G\) (formed by the edges for which \(\alpha_e = 0\)) does not contribute to the Landau equations; effectively, they turn into analogous equations for the \emph{reduced diagram}~\(G / \gamma\) obtained by contracting the subdiagram~\(\gamma\) to a point.
This means that solutions of the Landau equations in the degenerate case can be found by recursively solving the non-degenerate equations for the reduced diagrams.

A convenient way to implement this in a computer code is by using the following 
factorization formulae for the Symanzik polynomials (for example,
see ref.~\cite[Proposition~4.1]{Mizera:2021icv} and references therein):
\begin{subequations}
  \label{eq:UFfactor}
  \begin{align}
    \label{eq:Ufactor}
    \mathcal{U}_G \vert_{\alpha_{\gamma} \to \lambda \alpha_{\gamma}} &= 
    \lambda^{L_{\gamma}} \mathcal{U}_{\gamma} \, \mathcal{U}_{G / \gamma}
    + \mathcal{O} \left( \lambda^{L_{\gamma} + 1} \right), \\
    \label{eq:Ffactor}
    \mathcal{F}_G \vert_{\alpha_{\gamma} \to \lambda \alpha_{\gamma}} &= 
    \lambda^{L_{\gamma}} \mathcal{U}_{\gamma} \, \mathcal{F}_{G / \gamma}
    + \mathcal{O} \left( \lambda^{L_{\gamma} + 1} \right),
  \end{align}
\end{subequations}
where \(\alpha_{\gamma} \to \lambda \alpha_{\gamma}\) means that
\(\alpha_e\) are replaced by \(\lambda \alpha_e\) for all edges in \(\gamma\), 
while \(L_{\gamma}\) is the number of loops in the subdiagram~\(\gamma\).
\Cref{eq:Ufactor} also shows that \(\mathcal{U}\) vanishes only when \(\gamma\) 
contains at least one loop, so the degenerate case \(\mathcal{U} = 0\) 
actually corresponds to IR subdivergences with one or more loop momenta 
unconstrained.

The procedure just described for solving the Landau equations can be summarized as follows:
\begin{enumerate}
\item Compute the first Symanzik polynomial \(\mathcal{U}\) 
using~\cref{eq:Ucomp}.
\item Recursively find all reduced diagrams with one or more loops contracted 
to a point by solving the equation \(\mathcal{U} = 0\) monomial by monomial, and
using~\cref{eq:Ufactor} to evaluate \(\mathcal{U}\) for reduced diagrams.
\item For the original diagram and each reduced diagram:
  \begin{enumerate}
  \item compute the second Symanzik polynomial \(\mathcal{F}\) 
  using~\cref{eq:Fcomp,eq:Ffactor};
  \item if \(\mathcal{F}\) is subtraction-free, solve \(\mathcal{F} = 0\) 
  monomial by monomial, otherwise solve~\cref{eq:landau_par} explicitly, e.g., by finding a subtraction-free linear combination~\eqref{eq:landau_par_lin_comb};
  \item keep only solutions for which the corresponding \(\mathcal{U} \neq 0\).
  \end{enumerate}
\item For each solution, which is given as a set of equality constraints on 
\(\alpha_e\), evaluate the corresponding constraints on the loop momenta 
using~\cref{eq:landau_lin_sol}.
  These constraints describe the sought-after surfaces in the loop momentum 
  space which contain all IR divergences. 
\end{enumerate}

\subsection{IR Power Counting}
\label{sec:power_counting}

Solutions of the Landau equations turn out to impose two types of constraints 
on the loop momenta: \emph{soft} constraints take the 
form~\(q_e(\ell, k) = 0\), where \(q_e(\ell, k)\) is the momentum of a 
massless line, while \emph{collinear} constraints 
require \(q_e(\ell, k) = x k_i\), where \(k_i\) is an external massless 
momentum, and the proportionality coefficient \(x \in (0, 1)\) is the ratio of 
unconstrained \(\alpha\) parameters, 
such as \(\alpha_1 / (\alpha_1 + \alpha_2)\).

Following Anastasiou and Sterman~\cite{Anastasiou:2018rib}, we use the IR 
power-counting technique of Libby and 
Sterman~\cite{Sterman:1978bi,*Libby:1978qf,*Libby:1978bx} to determine the 
behavior of the integral near a divergent surface described by soft and 
collinear constraints.
Namely, we parametrize the vicinity of the divergent surface by modifying the 
constraints as follows:
\begin{equation}\label{eq:modified_constraints}
    \begin{split}
        q^\mu_e(\ell, k) &= 0
        \quad \;
        \quad \to \quad 
        q^\mu_e(\ell, k) = \lambda \sigma_e^\mu\,,
        \\
        q^\mu_e(\ell, k) &= x_e k_i^\mu
        \quad \to \quad 
        q^\mu_e(\ell, k) = x_e k_i^\mu + \lambda \eta_i^\mu+ \lambda^{1/2} q_e^{\perp,\mu}\,.
    \end{split}
\end{equation}
Here, \(\bar{q}_e\) is a unit euclidean-norm vector while 
\(\eta_i\) and \(q_e^{\perp}\) satisfy,
\begin{equation}\label{eq:light_cone_definition}
     q_e^{\perp} \cdot k_i = \eta^2_i = q_e^{\perp} \cdot \eta_i = 0
     \quad\text{and}\quad
     \hat \eta_i = 0\,.
\end{equation} 

Solving the modified set of constraints for the loop momenta, we obtain a 
scaling rule of the form \(\ell_i = \ell_i(\lambda)\), which can be used to 
find numerators canceling the divergence the same way as we did in the UV case.
For every divergent configuration of the loop momenta, we can 
substitute the scaling rules of \cref{eq:modified_constraints} into the integration measure, which yields, 
%
%
\begin{equation}
    \label{eq:measure_scalings}
\begin{split}
\text{soft:} \quad   \mathrm{d}^D \ell &= \mathrm{d} \Omega_{D} \:  \mathrm{d}\lambda \: \lambda^{D-1} \sim \mathrm{d}\lambda \: \lambda^{D-1} \, ,\\
\text{collinear:} \quad   \mathrm{d}^D \ell &= \frac{1}{2} \mathrm{d}x_e \: \mathrm{d} {q_e^\perp}^2 \:\mathrm{d} \Omega_{D-2} \: (k_i \cdot \eta_i) \: \mathrm{d}\lambda \: \lambda^{D/2-1} \sim \mathrm{d}\lambda \: \lambda^{D/2-1} \, ,
\end{split}
\end{equation}
as well as into the integrand obtained from the UV-compatible numerator 
obtained in the previous section. We then Laurent-expand the integrand 
in \(\lambda\) and retain only powers leading to divergences.
The coefficients of divergent powers of \(\lambda\) are in general 
polynomials in the quantities,
\begin{equation}
    \begin{split}
        & \ell_i \cdot v_j,\;\,
        \LLort{i}{j} ,\;\, 
        \hat{\ell}_i \cdot \hat{\sigma}_e ,\;\,
        \hat{\ell}_i \cdot \hat{q}^\perp_j,\;\,
        v_i \cdot q^\perp_j,\;\, 
        v_i \cdot \eta_j, \;\,
        \\&
        v_i \cdot \sigma_e,\;\,  
        \hat{\sigma}_e \cdot \hat{q}^\perp_j,\;\,
        \hat{q}^\perp_i \cdot \hat{q}^\perp_j,\;\, 
        \hat{\sigma}_{e_1} \cdot \hat{\sigma}_{e_2},\;\,
        x_e,    
    \end{split} \label{eq:monomials_limits}
\end{equation}
with coefficients depending on the ansatz parameters \(c_{\vec{r}}\), and \(\ell_i\) being the unconstrained loop momenta.  The set of monomials in \cref{eq:monomials_limits} is determined by substituting the modified constraints of \cref{eq:modified_constraints} into the ansatzes of \cref{eq:general_ansatz,eq:monomials_alt}. 

However, not all monomials in \cref{eq:monomials_limits} are independent. 
We can find relations by applying the decomposition of 
\cref{eq:vNV_decomposition} to each of the vectors appearing in 
\cref{eq:light_cone_definition}.
We find,
\begin{equation}\label{eq:extra_relations}
\sum_{j=1}^R (k_i \!\cdot\! k_j) \: v_j \!\cdot\! q_e^\perp \:  =
\sum_{j,h=1}^R (k_j \!\cdot\! k_h) \:v_j \!\cdot\! \eta_i \; v_h \!\cdot\! \eta_i \:= 
\sum_{j,h=1}^R (k_j \!\cdot\! k_h) \: v_j \!\cdot\! q_e^\perp \; v_h \!\cdot\! \eta_i \: =
0 \,,
\end{equation}
with $i$ labelling a massless external momentum $k_i$.
The first is a relation among degree-1 monomials while the other 
two are among degree-2 monomials.
We take these relations into account and require that the coefficient of every 
independent monomial should vanish. The solution of the associated system of 
equations leads us to a set of constraints on the ansatz parameters. 

The \(k_i \!\cdot\! k_j\) factors appearing in~\cref{eq:extra_relations} are the only 
kinematics-dependent objects which enter the system of linear equations 
induced by the IR finiteness constraints. Moreover, they are only relevant 
when power-like collinear singularities are present. In practice one can first 
solve the system of equations which does not involve these \(k_i \!\cdot\! k_j\) and then 
solve a much smaller system coming from the subleading contributions of the 
numerator which may possibly depend on the \(k_i \!\cdot\! k_j\). 
This greatly simplifies the determination of locally finite numerators 
and allows us to compute them for topologies up to four loops on a laptop. 
In practice when the problem involves 
large systems of equations we use \texttt{FiniteFlow}~\cite{Peraro:2019svx} to obtain a solution efficiently.

Repeating the procedure above for all IR-divergent surfaces, we find the most 
general numerator consistent with IR finiteness for a given ansatz. Denoting 
the space of locally finite numerators by $\mathcal{W}$, any element 
$f \in \mathcal{W}$ can then be written in the form,
\begin{equation} \label{eq:locally_finite_result}
    f(\ell) = \sum_i \, g_i \: \mathcal{N}_i(\ell)\,,
\end{equation}
where the $\mathcal{N}_i$ are polynomials in the independent monomials in 
\cref{eq:monomials_alt} and the $g_i$ are unconstrained coefficients 
independent of the loop momenta.

To guarantee IR finiteness, following ref.~\cite{Anastasiou:2018rib}, one 
must check not only the divergent surfaces found from the Landau analysis, 
but also subsurfaces.  The latter do not always show up as separate solutions 
of the Landau equations.
In particular, a solution with a soft constraint may be a special case of a solution with a collinear constraint when the proportionality coefficient~\(x_e\) reaches its boundary value of 0 or 1.
To include the subsurfaces we also consider each solution of the Landau equations where \(x_e\) have been set to 0 or 1 in all possible combinations.
This is necessary only when the degree of divergence
of a subsurface is greater than that of any of the parent surfaces. An example
of this would be a set of surfaces associated to logarithmic divergences
intersecting on a subsurface which leads to a power-like divergence, as in the
case of the massless non-planar double box, see \cref{sec:nonplanar_dbox}.

\subsection{Finite Integral Generators}

\newcommand{\WIR}{\mathcal{\widetilde W}}
Suppose we have the set of all locally IR-finite numerators 
$\WIR$, constructed as above without imposing UV finiteness.
Multiplying any locally IR-finite numerator $f \in \WIR$ 
by any polynomial $p$ in the variables of \cref{eq:monomials_alt} 
yields another locally IR-finite numerator:
\begin{equation}
    p f \in \WIR \, .
\end{equation}
This is the defining property of an \emph{ideal}; it implies that there exists a (non-unique) minimal set of polynomials $f_1,f_2,\dots \in \mathcal{W}$ which generate $\WIR$:
\begin{equation}
     \WIR = \langle f_1,f_2,\dots \rangle \, .
\end{equation}
This means that any IR-finite polynomial can be written as a linear combination of \(f_1, f_2, \dots\) with polynomial coefficients, and vice versa, any such combination is IR-finite.
As we shall see, only a small number of generators are required. 
This allows us to capture all information on $\WIR$ in a very compact form. 

Requiring UV finiteness then restricts us to a 
finite-dimensional subspace within the ideal.  We will call this
the \emph{truncated ideal\/} of locally finite numerators, though it is mathematically no longer an ideal.
We will order generators according to their UV behavior, 
favoring numerators with lower degrees in the loop momenta.  

One could of course impose the UV-finiteness and IR-finiteness conditions in either order; we will impose the UV-finiteness ones first.  The set of
generators we obtain will be the same.

\section{A Simple Example: One-Loop Box} 
\label{sec:onebox}
\begin{figure}[tb]
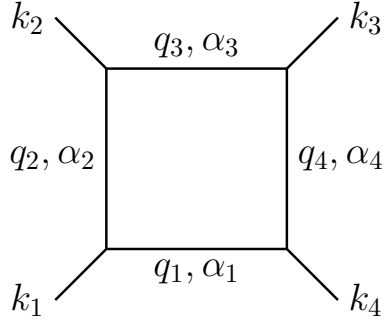

  \centering
    \SBox
  \caption{\label{fig:box}%
    The massless one-loop box graph.}
\end{figure}

To illustrate the proceedure described in the previous sections, 
let us consider the one-loop massless box integral, depicted in
\cref{fig:box}, as a simple example.  We define the momenta, 
\begin{equation}
    q_1 = \ell, \quad
    q_2 = \ell-k_1, \quad
    q_3 = \ell-k_{12}, \quad
    q_4 = \ell-k_{123} \,.
\end{equation}
We start by writing down the most general ansatz for a UV-finite numerator as 
described in \cref{sec:weinberg_uv}. Superficial convergence imposes a maximum 
degree $ r_{\mathrm{max}}=3 $ and as there are no subintegrations the ansatz reads,
\begin{equation} \label{eq:box_uv_num}
    \begin{split}
         \mathcal{N}(\ell) = \;
         &c_0 \, + \, c_1\:{\ell \cdot v_1}\, + \,  
        c_2\:{\ell \cdot v_2 }\, + \, 
        c_3\:{ \ell \cdot v_3}\, + \, 
        c_4\:{\Lort{}^2}\, + \, 
        c_5\:{(\ell \cdot v_1)^2 }\, + \, 
        c_6\:{\ell \cdot v_1 \:\ell \cdot v_2}
        \\&\! + \, 
        c_7\:{(\ell \cdot v_2)^2 }\, + \, 
        c_8\: {\ell \cdot v_1 \:\ell \cdot v_3} \, + \, 
        c_9\:{\ell \cdot v_2 \:\ell \cdot v_3} \, + \, 
        c_{10}\:{(\ell \cdot v_3)^2}  \, + \,  
        c_{11}\:{\Lort{}^2 \:\ell \cdot v_1}
        \\&\! + \, 
        c_{12}\:{\Lort{}^2 \:\ell \cdot v_2} \, + \, 
        c_{13}\:{\Lort{}^2 \:\ell \cdot v_3} \, + \, 
        c_{14}\:{(\ell \cdot v_1)^3} \, + \, 
        c_{15}\:{(\ell \cdot v_1)^2 \:\ell \cdot v_2 }
        \\&\! + \, 
        c_{16}\:{\ell \cdot v_1 \:(\ell \cdot v_2)^2 }\, + \, 
        c_{17}\:{(\ell \cdot v_2)^3 }\, + \, 
        c_{18}\:{(\ell \cdot v_1)^2 \:\ell \cdot v_3 }\, + \, 
        c_{19}\:{\ell \cdot v_1\: \ell \cdot v_2 \:  \ell \cdot v_3}
        \\&\!+ \, 
        c_{20}\:{(\ell \cdot v_2)^2 \:\ell \cdot v_3 }\, + \, 
        c_{21}\:{\ell \cdot v_1 \:(\ell \cdot v_3)^2 }\, + \, 
        c_{22}\:{\ell \cdot v_2 \:(\ell \cdot v_3)^2 }\, + \, 
        c_{23}\:{(\ell \cdot v_3)^3 }  \, .
    \end{split}
\end{equation}
Moving to the IR analysis, we start by computing
the auxiliary quantities (see \cref{eq:combined_den}),
\begin{equation}
\label{eq:box_AandB}
    \begin{split}
        &A_{11} = \alpha_1+\alpha_2+\alpha_3+\alpha_4 \,, \\
        &B_1^\mu= (\alpha_2+\alpha_3+\alpha_4) \: k_1^\mu + (\alpha_3+\alpha_4) \: k_2^\mu + \alpha_4 \: k_3^\mu \, ,
    \end{split}
\end{equation}
so that the Symanzik polynomials for this diagram are,
\begin{equation}
        \mathcal{U} = \alpha_1+\alpha_2+\alpha_3+\alpha_4 \,,  \qquad
        \mathcal{F} = s \, \alpha_1 \alpha_3 + t \, \alpha_2 \alpha_4 \, .
\end{equation}

Solving $\mathcal{F}\!=\!0$ monomial by monomial, and then using 
\cref{eq:landau_lin_sol} together with \cref{eq:box_AandB}, we obtain four solutions corresponding to divergent surfaces of collinear type.
Intersecting these surfaces, or equivalently setting the proportionality coefficients to 0 or 1, we find four soft subsurfaces. 
We list the results in \cref{tab:1loop_box_regions}.
\begin{table}[tb]
  \caption{\label{tab:1loop_box_regions}%
    Solutions of the Landau equations for the 1-loop box and the corresponding IR divergent surfaces and subsurfaces. 
    Teal dotted lines indicate collinear subdiagrams while red dashed lines represent soft propagators.}
  \centering
  \begin{tabular}{ccc @{\hspace{6em}} cc}
    surface & \makecell{solution\\of \(\mathcal{F} = 0\)} & \makecell{momentum\\constraint}
    & subsurface & \makecell{momentum\\constraint} \\
    \midrule
    \makecell{\SBoxRegionD\\$(C_1)$} & \(\alpha_3 = \alpha_4 = 0\) & \(q_1 = \frac{\alpha_2}{\alpha_1+\alpha_2} k_1\)
    & 
    \makecell{\SBoxRegionE\\$(S_1) = (C_4) \cap (C_1)$} & \(q_1 = 0\)
    \\
    \makecell{\SBoxRegionA\\$(C_2)$} & \(\alpha_4 = \alpha_1 = 0\) & \(q_2 = \frac{\alpha_3}{\alpha_2+\alpha_3} k_2\)
    &
    \makecell{\SBoxRegionH\\$(S_2) = (C_1) \cap (C_2)$} & \(q_2 = 0\)
    \\
    \makecell{\SBoxRegionB\\$(C_3)$} & \(\alpha_1 = \alpha_2 = 0\) & \(q_3 = \frac{\alpha_4}{\alpha_3+\alpha_4} k_3\)
    &
    \makecell{\SBoxRegionG\\$(S_3) = (C_2) \cap (C_3)$} & \(q_3 = 0\)  
    \\
    \makecell{\SBoxRegionC\\$(C_4)$} & \(\alpha_2 = \alpha_3 = 0\) & \(q_4 = \frac{\alpha_1}{\alpha_4+\alpha_1} k_4\)    
    & 
    \makecell{\SBoxRegionF\\$(S_4) = (C_3) \cap (C_4)$} & \(q_4 = 0\)
    \\
  \end{tabular}
\end{table}

\newcommand{\BoxName}{\mathrm{Box}}
To each of the singular configurations we associate a modified constraint 
$\ell = \ell(\lambda)$ as decribed in \cref{eq:modified_constraints} and 
substitute it into the momentum-space representation of the integral,
\begin{equation}
    \label{eq:box_momentum_rep}
    \BoxName[\mathcal{N}(\ell)] =
  \int  \mathrm{d}^D \ell \, \frac{\mathcal{N}(\ell)}{\mathcal{D}_1\, \mathcal{D}_2\, \mathcal{D}_3\, \mathcal{D}_4}\, .
\end{equation}
Using the integration-measure scalings~\eqref{eq:measure_scalings}
together with the scaling of the propagators $\mathcal{D}_e = q_e^2 + i \varepsilon$  we find that $ \BoxName[1] $ 
behaves as $\mathrm{d}\lambda \; \lambda^{-1 + \Ord(\epsilon)}$ near $D=4$ for 
all the soft and collinear configurations of \cref{tab:1loop_box_regions}. 
Because of the logarithmic nature of these singularities we need to
expand the numerator only to $\Ord(\lambda^0)$ to obtain singular contributions to the integral. To this order in $\lambda$ we find that the monomials $\{\Lort{}^2,\; \ell\cdot v_1,\; \ell\cdot v_2, \; \ell\cdot v_3\}$ take the values:
\begin{equation} \label{{box:monomial_scaling}}
    \begin{split}
        C_1: \;\{0,x_1,0,0\}\, , \quad
        C_2: \;\{0,1,x_2,0\}\, &, \quad
        C_3: \;\{0,1,1,x_3\}\, , \quad
        C_4: \;\{0,x_4,x_4,x_4\}\, , \\
        S_1: \;\{0,0,0,0\}\,, \quad
        S_2: \;\{0,1,0,0\}\, &, \quad
        S_3: \;\{0,1,1,0\}\, , \quad
        S_4: \;\{0,1,1,1\}\, ,
    \end{split}
\end{equation}
where the $x_e$ stand for the collinear-fraction parameters which can be 
read off from \cref{tab:1loop_box_regions} for the different collinear regions.

Let us study
these configurations,
extracting the corresponding constraints on the numerator. 
Substituting the collinear configurations into \cref{eq:box_uv_num}
and setting the result to zero for every value of the $x_e$ we find the equations,
\begin{equation}\label{eq:box_eq_collinear}
    \begin{split}
        C_1: \;& c_0 \:=\: c_1 \:=\: c_5 \:=\: c_{14} \:=\: 0 \, , \\
        C_2: \;& c_0 + c_{1} + c_{5} + c_{14} \:=\: c_{2} + c_{6} + c_{15}  \:=\:  c_{7} + c_{16}  \:=\:  c_{17} \:=\: 0 \, , \\
        C_3: \;& c_0 + c_{1} + c_{2} + c_{5} + c_{6} + c_{7} + c_{14} + \dots + c_{17} \:=\: \\
        \quad  & c_{3} + c_{8} + c_{9} + c_{18} + c_{19} + c_{20} \:=\: c_{10} + c_{21} + c_{22} \:=\: c_{23} \:=\: 0 \, , \\
        C_4: \;& c_0 \:=\: c_{1} + c_{2} + c_{3} \:=\: c_{5} + \dots  + c_{10} \:=\: c_{14} + \dots + c_{23} \:=\: 0\, .
    \end{split}
\end{equation}
As anticipated these equations do not depend on the external kinematic
invariants $s_{ij}$. Moving on, for each soft region $S_i$ we get the following constraints:
\begin{equation}\label{eq:box_eq_soft}
    \begin{split}
        S_1: \;&c_0 \:=\: 0 \, , \\
        S_2: \; &c_0 + c_1 + c_5 + c_{14} \:=\: 0 \, , \\
        S_3: \;&c_0 + c_1 + c_2 + c_5 + c_6 + c_7 + c_{14} + \dots + c_{17} \:=\: 0 \, , \\
        S_4: \; &c_0 + \dots  + c_3 + c_5 + \dots  + c_{10} + c_{14} + \dots  + c_{23} \:=\: 0 \, .
    \end{split}
\end{equation}
One can check that these constraints are satisfied automatically if \cref{eq:box_eq_collinear} holds.
This comes as no surprise, because in this case soft subsurfaces have the same degree of divergence as their parent surfaces corresponding to collinear regions (see discussion at the end of \cref{sec:power_counting}).
Therefore, cancellation of collinear divergences is sufficient for IR-finiteness.

The constraints in \cref{eq:box_eq_collinear,eq:box_eq_soft} entirely fix
12 coefficients in the ansatz of \cref{eq:box_uv_num}. Therefore, within the space of UV-finite numerators of \cref{eq:box_uv_num}, described by 24 independent monomials, the subset of IR-finite integrals has 12 degrees of freedom.

In general, the results of a UV--IR analysis such as the one presented
above can be summarized by simply listing the generators of the 
ideal of IR-finite numerators.  A set of generators can be found
using standard techniques of computational algebraic geometry,
such as Gr{\"o}bner bases, and then prettifying the results.
In the case of the massless one-loop box 
there are only three rank-two generators:  $\Lort{}^2$,  $( \ell-k_1)\cdot v_1 \: \ell \cdot (v_2-v_3)$ and  $ \ell\cdot v_3 \: \ell \cdot ( v_1 - v_2)$.
Using them we can write the most general UV- and IR-finite numerator for the massless one-loop box as,
\begin{equation} \label{eq:box_finite_numerator}
    \begin{split}
    \mathcal{N}(\ell) = \; 
    &[ b_1 \:+\: b_2 \,\ell \cdot v_1 \:+\: b_3 \,\ell \cdot v_2 \:+\: b_4 \,\ell \cdot v_3 ] \; ( \ell-k_1)\cdot v_1 \: \ell \cdot (v_2-v_3) \; + \\
    &[ b_5 \:+\: b_{6} \,\ell \cdot v_1 \:+\: b_{7} \,\ell \cdot v_2 \:+\: b_{8} \,\ell \cdot v_3 ] \;  \ell\cdot v_3 \: \ell \cdot ( v_1 - v_2) \; + \\
    & [ b_9 \:+\: b_{10} \,\ell \cdot v_1 \:+\: b_{11} \,\ell \cdot v_2 \:+\: b_{12} \,\ell \cdot v_3 ] \; \Lort{}^2 \, ,
    \end{split}
    \end{equation}
where the $b_i$ are purely functions of the external 
momenta. 

We end our discussion with two comments on \cref{eq:box_finite_numerator}:
\begin{enumerate}
    \item The generators obtained in the van Neerven--Vermaseren basis of monomials have a straightforward representation in terms of Gram determinants, 
    \begin{equation}
        \begin{split}
             ( \ell-k_1)\cdot v_1 \: \ell \cdot (v_2-v_3) &\propto \Gram{\ell-k_1 & 2 & 3}{1 & 2 & 3} \Gram{\ell & 1 & 4}{1 & 2 & 3}\, , \\
              \ell\cdot v_3 \: \ell \cdot ( v_1 - v_2) &\propto \Gram{\ell & 1 & 2}{1 & 2 & 3} \Gram{\ell & 3 & 4}{1 & 2 & 3} \, ,\\
              \Lort{}^2 & \propto \GramO{\ell & 1 & 2 & 3}\, ,
        \end{split}
    \end{equation}
    where the constants of proportionality depend only on the external
    momenta (here and in what follows we represent the external momenta~\(k_i\) inside the Gram determinants with the corresponding labels~\(i\)). This observation generalizes to higher loops.    
    \item \Cref{eq:box_finite_numerator} is just one possible choice to represent the same numerator. For instance, one could swap the 
    $\ell \cdot v_i$ factors inside the brackets in
    \cref{eq:box_finite_numerator} for  
    $\ell \cdot k_i$ or rewrite them in terms of inverse propagators where 
    possible. As long as the UV power-counting is satisfied one can choose different representations depending on the context.
\end{enumerate}

\section{Evanescent Integrands}%
\label{sec:evanescent}

Locally finite integrals, enumerated by the algorithmic procedure described 
above, contain a noteworthy subset of integrals which are manifestly of $\Ord(\eps)$ while containing no explicit appearances of $\eps$ (or $D$). 
We call these \emph{evanescent\/} integrals.
To find integrands giving rise to evanescent integrals, we start with the 
set of integrands of locally finite integrals, and impose further 
restrictions on their coefficients.  
In particular we can split the numerator of the integrand,
\begin{equation}
    \mathcal{N}= \mathcal{N}_4 + \mathcal{N}_{D-4}\, ,
\end{equation} 
where $\mathcal{N}_4 = \mathcal{N}|_{D=4}$, so that $\mathcal{N}_{D-4}$ 
vanishes exactly in four dimensions for any value of the loop momenta.
If the integral $\mathcal{I}[\mathcal{N}]$ is finite then $\mathcal{I}[\mathcal{N}_4]$ is 
also finite as our procedure to determine locally finite integrands could in
principle be performed purely in four dimensions. 
Rewriting the equation above as,
\begin{equation}
   \mathcal{N}_{D-4} =  \mathcal{N}- \mathcal{N}_4 \,,
\end{equation} 
we see that as $D \to 4$ the right-hand side has to vanish and therefore $\mathcal{N}_{D-4}$ can be at most of $\Ord(\eps)$.
To find evanescent integrands we therefore require the integrand to vanish when all loop momenta are four-dimensional, that is
$\mathcal{N}_4 = 0$.

For $(n > 4)$-point integrals, setting the $(D\!-\!4)$-dimensional components of each loop momentum to zero corresponds to fixing the extra components $\Lort{i}=0$ so that all $\LLort{i}{j}$ vanish identically. For the numerator to vanish as $D\to 4$ we require a vanishing coefficient 
for every monomial of the independent scalar products 
$\ell_i \cdot v_{j=1,\cdots,4}$. 
The truncated ideal of evanescent numerators 
is then the intersection of the 
locally finite truncated ideal $\mathcal{W}$ 
with the one generated by the $\LLort{i}{j}$.

For $(n \leq 4)$-point integrals we need to be more careful: we can still freely set to zero the coefficient of every monomial without a factor of $\LLort{i}{j}$, as these are linearly independent and do not vanish as $D\to4$. Conversely $\LLort{i}{j} \neq 0$ in four space-time dimensions, so evanescence of a term containing one of these factors is not guaranteed. Instead the $\Lort{i}$ become linearly dependent if enough loops are present because they span a $(5-n)$-dimensional space. As a consequence, any Gram determinant with $m \geq 6-n$ entries of the $\Lort{i}$'s will automatically vanish. For instance, with $n=4$  we need at least two loops to find an evanescent numerator via the factor 
\begin{equation}
    \GramO{\Lort{1} & \Lort{2} } \, .
\end{equation}
This is in accordance with the fact that for the one-loop box no evanescent numerator can be extracted from \cref{eq:box_finite_numerator}.
In general, within the space of polynomials defined by \cref{eq:general_ansatz}, the ideal of evanescent numerators for $n \leq 4$ is given by 
the intersection of the locally finite ideal with the ideal generated by all choices of the Gram determinant,
\begin{equation}
    \Gram{\Lort{i_1} & \dots & \Lort{i_{6-n}}}{\Lort{j_1} & \dots & \Lort{j_{6-n}}} \, .
\end{equation}

\paragraph*{Evanescently-Finite Integrals} 
Integrands can vanish in four dimensions even if they fail the
UV-finiteness power-counting criterion, or if they fail to cancel all
IR divergences revealed by the Landau equations.  In dimensional regularization,
these combinations can give rise to \emph{evanescently finite} integrals.
These are integrals which are finite, but where the finiteness depends
on cancellation of a pole in $\eps$ (due to a UV or IR divergence) with a factor of $\eps$ that arises from
the evanescence properties of the integrand.  Their existence is special
to dimensional regularization. 
We will not study them exhaustively, but provide explicit examples in \cref{sec:results}.

\subsection{Massless Pentagon}
\begin{figure}[tb]
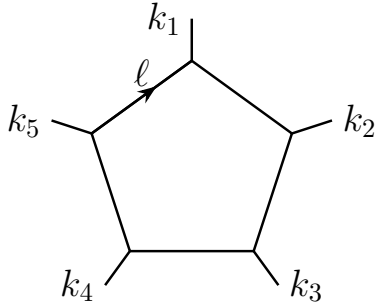

    \centering
    \Pentagon
    \caption{\label{fig:pentagon}%
      Massless Pentagon}
\end{figure}

The massless pentagon, depicted in \cref{fig:pentagon}, is the simplest one-loop case where evanescent integrals arise. 
Let us summarize the results of our procedure applied to the pentagon. 
We start by writing a UV-compatible ansatz for the numerator, here
a rank-five polynomial in the variables,
\begin{equation}
    \{{\Lort{}}^2,\; \ell\cdot v_1, \; \ell\cdot v_2, \; \ell\cdot v_3, \; \ell\cdot v_4\} \, .
\end{equation}
This polynomial has 166 independent monomials whose coefficients are unfixed rational functions of the independent Mandelstam variables, for instance the set \( \{s_{12},s_{23},s_{34},s_{45},s_{51}\}\).  

We proceed by listing all IR singularities. 
The list generalizes that for the box integral (see \cref{tab:1loop_box_regions}): 
here we find a total of ten regions 
(five soft, five collinear). Requiring the numerator to vanish
appropriately in 
each of the IR regions, we obtain a set of 141 locally finite 
independent numerators.  The ideal needs only six generators:\\
\begin{equation}
\label{eq:pentagon_generators}
\begin{split}
    & 
    \ell\cdot v_4 \; \ell \cdot (v_1-v_2)\,, 
    \quad 
   (\ell -k_1)\cdot v_1 \; \ell \cdot (v_2-v_3)\,, 
    \quad 
    \ell \cdot (v_1-v_2) \; \ell\cdot (v_3-v_4) \,, 
    \\
    &
    \ell \cdot (v_2-v_3) \; \ell\cdot v_4 \,, 
    \quad 
    \ell \cdot (v_3-v_4) \; (\ell - k_1)\cdot v_1 \,, 
    \quad
    {\Lort{}}^2 \, .
\end{split}
\end{equation}
The same ideal can equivalently be expressed in terms of the closely related set of Gram determinants:
\begin{equation}
\label{eq:pentagon_generators_grams}
\begin{split}
    &\Gram{\ell & 1 & 2 & 3}{\ell & 3 & 4 & 5}\,, \quad 
    \Gram{\ell - k_1 & 2 & 3 & 4}{\ell & 4 & 5 & 1}\,, \quad 
    \Gram{\ell & 3 & 4 & 5}{\ell & 5 & 1 & 2}\,, \\[6pt]
    &\Gram{\ell & 4 & 5 & 1}{\ell & 1 & 2 & 3}\,, \quad 
    \Gram{\ell & 5 & 1 & 2}{\ell - k_1 & 2 & 3 & 4}\,, \quad 
     \GramO{\ell & 1 & 2 & 3 & 4} \, .
\end{split}
\end{equation}
Turning to evanescent integrands, the further requirement that the numerator should vanish when the loop momenta are 
purely four-dimensional leads us to a subset of 40 evanescent integrals, 
whose numerators belong to the ideal generated by ${\Lort{}}^2$ 
(\( =\GramO{\ell & 1 & 2 & 3 & 4}/\GramO{1 & 2 & 3 & 4} \)), which is consistent with the well-known result,
\begin{equation}
    \int d^D l \frac{\GramO{\ell & 1 & 2 & 3 & 4} }{ \mathcal{D}_1 \cdots \mathcal{D}_5} \propto (D-4) \times 
    \int d^{D+2} l \frac{\GramO{1 & 2 & 3 & 4} }{ \mathcal{D}_1 \cdots \mathcal{D}_5} = \mathcal{O}(\eps)\,.
\end{equation}

\section{Results}%
\label{sec:results}
In this section we present some explicit results. First,
we apply our proceedure to
characterize the locally finite integrals for the planar
double box, and then show how a general conjecture for the all-loop ladder 
topology can be justified.
We then move to consider 
the
non-planar double box and another two-loop four-point
graph which we refer to as the beetle graph. Both the non-planar double box and
the beetle require dealing with powerlike
IR singularities and allow us to illustrate aspects
of our procedure.

\subsection{Planar Double Box}%
\label{sec:planar_dbox}

In this section, we apply the procedures 
of \cref{sec:IR,sec:weinberg_uv} to the massless planar double box integral.
\begin{figure}[tb]
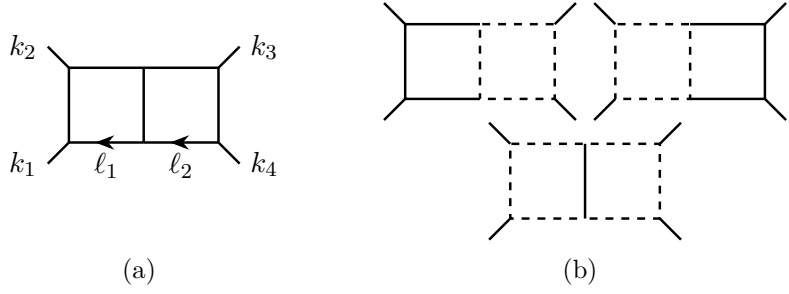

  \centering
  \newcommand{\SubfigureB}{
      \(
      \setlength{\extrarowheight}{5ex}
      \begin{array}[t]{cc}
        \DBoxBeforeA & \DBoxBeforeB \\
        \multicolumn{2}{c}{\DBoxBeforeC} 
      \end{array}
      \)
  }
  \subfloat[]{%
    \label{fig:planar_dbox_graph}
    \vphantom{\SubfigureB}\DBox
  }
  \qquad
  \subfloat[]{%
    \label{fig:planar_dbox_reductions}
    \SubfigureB
  }
  \caption{\label{fig:planar_dbox}%
    The planar double box graph (a) and and its sub-integrations (b).}
\end{figure}
We parametrize the loop momenta as shown in \cref{fig:planar_dbox_graph}.
All external momenta are outgoing and we label the edges from 1 to 7,
so that their momenta read, in order of labels,
\begin{equation}
  \ell_1,\quad
  \ell_1 - k_1,\quad
  \ell_1 - k_{12},\quad
  \ell_2,\quad
  \ell_2 - k_{123},\quad
  \ell_2 - k_{12},\quad
  \ell_1 - \ell_2.
\end{equation}
We start by imposing UV-convergence constraints and then proceed to the 
IR analysis described in previous sections.

\subsubsection{UV divergences}
To avoid the overall UV divergence, the maximal allowed numerator rank 
is 5, in accordance with \cref{eq:UV_rank_bound}. 
Three distinct linear combinations of the loop momenta enter the 
denominators: \(\ell_1\), \(\ell_2\), and \((\ell_1 - \ell_2)\).
Holding each one of them fixed in turn yields three possible subintegrations 
over the remaining variable~\(\ell\) corresponding to the dashed lines 
in \cref{fig:planar_dbox_reductions}. These give rise to the following constraints on the numerator:
\begin{equation}
 \lim_{\rho \to \infty} 
\begin{cases}
 \rho^{-4} \; \mathcal{N}\left( \ell_1 = \mathrm{const} , \; \ell_2 = \rho \ell \right) = 0\,, \\
 \rho^{-4} \; \mathcal{N}\left( \ell_1 = \rho \ell, \; \ell_2 = \mathrm{const} \right) = 0\,, \\
 \rho^{-8} \; \mathcal{N}\left( \ell_1 = \rho \ell, \; \ell_2 = \mathrm{const} + \rho \ell \right) = 0\,. 
\end{cases}
\end{equation}
The first and second constraints simply count the powers of  
\(\ell_2\) and \(\ell_1\) respectively, so the numerator has to be at most 
cubic in either of the loop momenta. The third constraint is automatically satisfied for numerators of rank five.
Hence, a numerator for the double box yields a UV-finite integral if it is at most cubic in~\(\ell_1\), at most cubic in~\(\ell_2\), and has a total rank of at most five.

\subsubsection{IR divergences}

Following our general algorithm of~\cref{sec:landau_solve}, we start by identifying all reduced diagrams of the double box. We do this by 
computing the first Symanzik polynomial:
\begin{equation}
  \mathcal{U} = (\alpha_1 + \alpha_2 + \alpha_3) (\alpha_4 + \alpha_5 + \alpha_6) + \alpha_7 (\alpha_1 + \alpha_2 + \alpha_3 + \alpha_4 + \alpha_5 + \alpha_6) \,,
\end{equation}
and setting \(\mathcal{U} = 0\) term by term. This yields three solutions 
which we collect in \cref{tab:planar_dbox_degenerate}.
\begingroup
\renewcommand{\arraystretch}{\mysize}
\begin{table}[tb]
  \caption{\label{tab:planar_dbox_degenerate}%
    One-loop reduced diagrams of the planar double box and corresponding Symanzik polynomials.}
  \centering
  \begin{tabular}{lcl}
   \makecell[c]{reduced graph} & \makecell[c]{solution of $\mathcal{U}=0$\\(full graph)} & \makecell[c]{Symanzik polynomials\\(reduced graph)} \\
    \midrule
    \addlinespace
    \DBoxBeforeA $\to$ \DBoxAfterA & \(\alpha_4 = \alpha_5 = \alpha_6 = \alpha_7 = 0\) & \(\begin{aligned} \mathcal{U} &= \alpha_1 + \alpha_2 + \alpha_3 \\ \mathcal{F} &= s \, \alpha_1 \alpha_3 \end{aligned}\) \\
    \addlinespace
    \DBoxBeforeB $\to$ \DBoxAfterB & \(\alpha_1 = \alpha_2 = \alpha_3 = \alpha_7 = 0\) & \(\begin{aligned} \mathcal{U} &= \alpha_4 + \alpha_5 + \alpha_6 \\ \mathcal{F} &= s \, \alpha_4 \alpha_6 \end{aligned}\) \\
    \addlinespace
    \DBoxBeforeC $\to$ \DBoxAfterC & \(\alpha_1 = \alpha_2 = \alpha_3 = \alpha_4 = \alpha_5 = \alpha_6 = 0\) & \(\begin{aligned} \mathcal{U} &= \alpha_7 \\ \mathcal{F} &= 0 \end{aligned}\)
  \end{tabular}
\end{table}
\endgroup
These solutions describe the reduced diagrams of~\cref{fig:planar_dbox_reductions}.
Because these diagrams already have only one loop, they represent all reduced diagrams of interest for solving the degenerate-case Landau equations: indeed, solving \(\mathcal{U} = 0\) monomial by monomial for the reduced diagrams 
gives no further solutions with at least one \(\alpha_i\) positive.

We now have to solve the parameter-space Landau equations for the original diagram and the three reduced diagrams.
We start with the planar double box diagram itself.
As the graph is planar, it is natural to express the second Symanzik polynomial~\(\mathcal{F}\) in terms of the independent Mandelstam invariants of consecutive external momenta, \(s = (k_1 + k_2)^2\) and \(t = (k_2 + k_3)^2\), so that \(\mathcal{F}\) is subtraction-free:
\begin{equation}
  \mathcal{F} = s \left[ \alpha_1 \alpha_3 (\alpha_4 + \alpha_5 + \alpha_6) + \alpha_4 \alpha_6 (\alpha_1 + \alpha_2 + \alpha_3) + \alpha_7 (\alpha_1 + \alpha_4) (\alpha_3 + \alpha_6) \right] + t \, \alpha_2 \alpha_5 \alpha_7\,.
\end{equation}
We find ten solutions of \(\mathcal{F} = 0\) monomial by monomial.  They
are listed in the first column of \cref{tab:planar_dbox_nondegenerate}.
\begingroup
\renewcommand{\arraystretch}{\mysize}
\begin{table}[tb]
  \caption{\label{tab:planar_dbox_nondegenerate}%
    Non-degenerate solutions of the Landau equations for the planar double box.}
  \centering
  \begin{tabular}{l @{\hspace{3em}} ll}
    solution of \(\mathcal{F} = 0\) & \multicolumn{2}{c}{ momentum constraints } \\
    \midrule
    \(\alpha_3 = \alpha_5 = \alpha_6 = 0\) & \(\ell_1 \parallel k_1 \) & \( \ell_2 \parallel k_1\) \\
    \(\alpha_2 = \alpha_3 = \alpha_6 = 0\) & \(\ell_1 \parallel k_4 \) & \( \ell_2 \parallel k_4\) \\
    \(\alpha_1 = \alpha_4 = \alpha_5 = 0\) & \((\ell_1 - k_{12}) \parallel k_2 \) & \( (\ell_2 - k_{12}) \parallel k_2\) \\
    \(\alpha_1 = \alpha_2 = \alpha_4 = 0\) & \((\ell_1 - k_{12}) \parallel k_3 \) & \( (\ell_2 - k_{12}) \parallel k_3\) \\
    \(\alpha_3 = \alpha_6 = \alpha_7 = 0\) & \(\ell_1 \parallel k_1 \) & \( \ell_2 \parallel k_4\) \\
    \(\alpha_3 = \alpha_4 = \alpha_7 = 0\) & \(\ell_1 \parallel k_1 \) & \( (\ell_2 - k_{12}) \parallel k_3\) \\
    \(\alpha_1 = \alpha_6 = \alpha_7 = 0\) & \((\ell_1 - k_{12}) \parallel k_2 \) & \( \ell_2 \parallel k_4\)  \\
    \(\alpha_1 = \alpha_4 = \alpha_7 = 0\) & \((\ell_1 - k_{12}) \parallel k_2 \) & \( (\ell_2 - k_{12}) \parallel k_3\) \\
    \(\alpha_1 = \alpha_2 = \alpha_3 = \alpha_7 = 0\) & \multicolumn{2}{c}{---} \\
    \(\alpha_4 = \alpha_5 = \alpha_6 = \alpha_7 = 0\) & \multicolumn{2}{c}{---} \\
  \end{tabular}
\end{table}
\endgroup
The last two solutions should be discarded as they nullify \(\mathcal{U}\) as well.
For the remaining eight solutions, we evaluate the corresponding loop momenta, 
recognizing the well-known double-collinear IR divergences of the planar double box.  For example, the solution on the first line of \cref{tab:planar_dbox_nondegenerate} gives,
\begin{equation}
\label{eq:solution_example_dbox}
    \ell_1
    = \frac{\alpha_2 (\alpha_4 + \alpha_7)}{(\alpha_1 + \alpha_2)(\alpha_4 + \alpha_7) + \alpha_4 \alpha_7} \, k_1 \,,
    \qquad
    \ell_2
    = \frac{\alpha_2 \alpha_7}{(\alpha_1 + \alpha_2)(\alpha_4 + \alpha_7) + \alpha_4 \alpha_7} \, k_1 \,,
\end{equation}
describing a double-collinear IR divergence \((\ell_1 \parallel k_1, \, \ell_2 \parallel k_1)\).

Computing all intersections of the surfaces associated with the singular configurations in \cref{tab:planar_dbox_degenerate}, or equivalently setting collinearity coefficients to zero or one, we reproduce the double-soft and soft-collinear IR subdivergences. For example,
intersecting \((\ell_1 \parallel k_1, \, \ell_2 \parallel k_1)\) with \((\ell_1 \parallel k_1, \, \ell_2 \parallel k_4)\) gives rise to the soft-collinear configuration \((\ell_1 \parallel k_1, \, \ell_2 = 0)\) as it sets $\alpha_7=0$ in \cref{eq:solution_example_dbox}.
Similarly, intersecting \((\ell_1 \parallel k_1, \, \ell_2 \parallel k_1)\) with \((\ell_1 \parallel k_4, \, \ell_2 \parallel k_4)\) yields the double-soft divergence \((\ell_1=0 , \, \ell_2 = 0)\) setting $\alpha_2=0$ in \cref{eq:solution_example_dbox}.

Notice that \(\ell_1 = 0\) implies \(\alpha_2 \alpha_7 = 0\). This forbids the configuration \((\ell_1 =0 , \, \ell_2 \parallel k_1)\) corresponding to a disconnected collinear diagram. More generally the eight solutions of \cref{tab:planar_dbox_nondegenerate} give rise to subdivergences which are in agreement with the results presented in ref.~\cite{Anastasiou:2018rib}.

Let us now turn to the reduced graphs of \cref{tab:planar_dbox_degenerate}. 
We repeat the same procedure: find all solutions of $\mathcal{F}=0$
monomial by monomial, and then compute the corresponding constraints on the loop momenta. The results of the reduced-graph analysis are given in \cref{tab:planar_dbox_degenerate_solutions}.
The first two graphs produce the familiar single-collinear divergences, as well as the single-soft divergences from intersections of solutions.
The last graph gives rise to an additional single-soft configuration, which is integrable by power-counting even with a trivial numerator, therefore we discard it (if we allowed higher powers of denominators, it could in principle become divergent).
\begingroup
\renewcommand{\arraystretch}{\mysize}
\begin{table}[tb]
  \caption{\label{tab:planar_dbox_degenerate_solutions}%
    Solutions of the Landau equations for the one-loop reduced diagrams of the planar double box.}
  \centering
  \begin{tabular}{ccl}
    reduced graph & solutions of \(\mathcal{F} = 0\) & momentum constraint \\
    \midrule
    \addlinespace
    \DBoxAfterA & \makecell{\(\alpha_3 = 0\)\\ \(\alpha_1 = 0\)} & \makecell[l]{\(\ell_1 \parallel k_1\)\\ \((\ell_1 - k_{12}) \parallel k_2\)} \\
    \addlinespace
    \DBoxAfterB & \makecell{\(\alpha_6 = 0\)\\ \(\alpha_4 = 0\)} & \makecell[l]{\(\ell_2 \parallel k_4\)\\ \((\ell_2 - k_{12}) \parallel k_3\)} \\
    \addlinespace[2ex]
    \DBoxAfterC & \(\alpha_7 \neq 0\) & \(\ell_1 - \ell_2 = 0\) \\
  \end{tabular}
\end{table}
\endgroup

As we now have the full set of singular configurations for the loop momenta, we can proceed by imposing finiteness constraints on the UV-finite numerator ansatz, as described in \cref{sec:power_counting}.
Because all divergences (including subdivergences) are logarithmic, it suffices to require that the numerator vanishes on all collinear configurations, as in the one-loop box case.
In \cref{tab:dbox_count} we list the numbers of linearly independent finite and 
evanescent ($\Ord(\eps)$) integrals up to a given rank, along with the number 
of the corresponding generators arising at each rank.
We construct them in the next section. 
\begingroup
\renewcommand{\arraystretch}{\mysize}
\begin{table}[tb]
  \caption{\label{tab:dbox_count}%
    Results of the double box numerator analysis.}
  \centering
  \begin{tabular}{lccccc}
    rank & 1 & 2 & 3 & 4 & 5 \\
    \midrule
    \# finite integrals & 0 & 2 & 18 & 89 & 247 \\
    \# finite generators & 0 & 2 & 4 & 4 & 0 \\
    \# evanescent integrals & 0 & 0 & 0 & 1 & 7 \\
    \# evanescent generators & 0 & 0 & 0 & 1 & 0 \\
  \end{tabular}
\end{table}
\endgroup

\subsubsection{Finding a Basis of Generators}

Thus far, we have identified the truncated ideals corresponding to 
locally finite and
evanescent numerators for the double-box integral.  We now turn to finding a compact and nice representation 
for the generators of these ideals. 
To do so we start by defining\footnote{These definitions depend both on the loop-momentum 
routing and the choice made for momentum conservation ($k_4 = -k_1-k_2-k_3$ in our case). 
Any changes will be reflected on the explicit form of the~$\FlowSymb{}$s.} 
the rank-one monomials,
\begin{equation}%
  \label{eq:ladder_building blocks_2loop}
  \begin{gathered}   
    \FlowSymb_1 = \ell_1 \! \cdot \!v_2\,, \quad 
    \FlowSymb_2 = (\ell_1 -k_{12}) \! \cdot \!v_1\, ,\quad 
    \FlowSymb_3 = (\ell_2-k_{12}) \! \cdot \!v_2\,, \quad 
    \FlowSymb_4 = \ell_2 \! \cdot \!(v_2-v_3)\, , 
    \\ 
    \FlowSymb_{12} = \ell_1 \! \cdot \!v_3\, , \quad
    \FlowSymb_{34} = (\ell_2- k_{123}) \! \cdot \!(v_1 - v_2)\, .
  \end{gathered}
\end{equation}
These turn out to have simple representations in terms of Gram determinants 
(see \cref{sec:doublebox_gram_ideal}). 
The $\FlowSymb_i$ variables are rank-one numerators which vanish in 
the collinear region associated to the $k_i$; 
the $\FlowSymb_{ij}$ are rank one
which vanish in the collinear regions associated to both $k_i$ and $k_j$. 
In addition, $\FlowSymb_1\FlowSymb_2$ and $\FlowSymb_{12}$ vanish in all soft limits involving $\ell_1$; by symmetry, $\FlowSymb_3\FlowSymb_4$ and $\FlowSymb_{34}$  vanish in those involving $\ell_2$.

Thanks to their properties we can use the $\FlowSymb$s together with the $\LLort{i}{j}$ as building blocks for the basis of the ideal of locally finite numerators for the double box.
We write down the simplest combinations that cancel all divergences, making sure not 
to repeat any lower rank basis element when writing higher rank ones.
Starting at rank two, we find that only the combinations 
$\FlowSymb_{12} \: \FlowSymb_{34}$ and $\LLort{1}{2}$ are locally finite. 
In fact, they correspond to the desired rank-two generators. 
At rank three we have more options. 
For instance, we can tame the singularities associated with legs 1 and 2 using two different $\FlowSymb$s,
and the ones of legs 3 and 4 with a single one: 
$\FlowSymb_1 \: \FlowSymb_2 \: \FlowSymb_{34}$. 
By symmetry we can also choose 
$\FlowSymb_{12} \: \FlowSymb_3 \: \FlowSymb_4$. In addition, because the $\Lort{i}$ vanish in any IR limit, we can write down two more rank-three
generators, $\LLort{1}{1}\: \FlowSymb_{34} $ and $\LLort{2}{2}\: \FlowSymb_{12}$. 
Finally, at rank four we have the following options:
\begin{enumerate}
\item cancel the divergence on each corner separately using the
product $\FlowSymb_1 \: \FlowSymb_2 \: \FlowSymb_3 \: \FlowSymb_4$;
\item use $\FlowSymb_1$ and $\FlowSymb_2$ to remove the singularities associated 
with $\ell_1$, and $\LLort{2}{2}$ to remove those of $\ell_2$, using the
product  $\FlowSymb_1 \: \FlowSymb_2 \: \LLort{2}{2}$;
\item use the flipped version $\FlowSymb_3\: \FlowSymb_4\: \LLort{1}{1}$;
\item use only the $\Lort{}$ components of the loop momenta, with
the product $\LLort{1}{1} \:\LLort{2}{2}$.
\end{enumerate}
Together, these options give us all four rank-four generators.
Summarizing, we find the double box generators,
\begin{equation}\label{eq:double_box_generators_results}
    \begin{aligned}
        &\text{rank\ two}: &&\quad 
        \FlowSymb_{12} \:\FlowSymb_{34}\, , \quad 
        \LLort{1}{2}\, ,
        \\
        &\text{rank three}: &&\quad 
        \FlowSymb_{1} \:\FlowSymb_{2} \:\FlowSymb_{34}\, , \quad 
        \FlowSymb_{12} \:\FlowSymb_{3} \:\FlowSymb_{4}\, , \quad
        \LLort{1}{1}\:\FlowSymb_{34}\, , \quad
        \LLort{2}{2}  \: \FlowSymb_{12}\, ,
        \\
        &\text{rank four}: &&\quad 
        \FlowSymb_{1} \:\FlowSymb_{2} \:\FlowSymb_{3} \:\FlowSymb_{4}\, , \quad 
        \LLort{1}{1} \:\FlowSymb_{3} \:\FlowSymb_{4}\, , \quad
        \LLort{2}{2} \:\FlowSymb_{1} \:\FlowSymb_{2}\, , \quad
        \LLort{1}{1} \: \LLort{2}{2} \, .
    \end{aligned}
\end{equation}
Using computational algebraic geometry, we have checked that this set of generators is non-redundant and indeed generates the entire truncated ideal of locally finite numerators.

Because we have only three independent external momenta, 
we can look to Gram determinants built out of the $\Lort{}$ components of the available 
loop momenta to build evanescent generators ($\Ord(\epsilon)$). At two loops the 
only possible combination is,
\begin{equation}
 \GramO{\Lort{1} & \Lort{2}} = \LLort{1}{1} \: \LLort{2}{2}- \LLort{1}{2}^2 \propto \GramO{\ell_1&\ell_2&k_1&k_2&k_3} \,.
\end{equation}
This is indeed the only evanescent generator, as can be seen from the 
counting in \cref{tab:dbox_count}. The evanescent ideal is then generated by, 
\begin{equation}\label{eq:double_box_generators_results_evanescent}
    \begin{split}
        \LLort{1}{1}\:\LLort{2}{2}- \LLort{1}{2}^2 \, .
    \end{split}
\end{equation}
Every evanescent numerator is by definition also locally finite, and this generator
can be written in terms of the generators in 
\cref{eq:double_box_generators_results}.
In order to make the subset of evanescent numerators more manifest within
the locally finite ones, we could alternatively choose the rank-four
generators for the planar double box to be,
\begin{equation}\label{eq:double_box_generators_results2}
    \begin{aligned}
        &\quad 
        \FlowSymb_{1} \:\FlowSymb_{2} \:\FlowSymb_{3} \:\FlowSymb_{4}\, , \quad 
        \LLort{1}{1} \:\FlowSymb_{3} \:\FlowSymb_{4}\, , \quad
        \LLort{2}{2} \:\FlowSymb_{1} \:\FlowSymb_{2}\, , \quad
        \LLort{1}{1}\:\LLort{2}{2}- \LLort{1}{2}^2 \, .
    \end{aligned}
\end{equation}

Beyond evanescence, we can also form combinations of locally finite
integrands that for symmetry
reasons give rise to identically vanishing integrals (\textit{i.e.} to all orders in \(\eps\)).  
We will not discuss these any further.  We also leave the question
of combining local finiteness with integration-by-parts reduction to
future investigation.\\

We can obtain examples of evanescently-finite integrands for the planar double 
box within the evanescent ideal by selecting UV-divergent numerators such as $(\LLort{1}{1} \: \LLort{2}{2}- \LLort{1}{2}^2)\: \ell_1 \cdot \ell_2$ or $(\LLort{1}{1} \: \LLort{2}{2}- \LLort{1}{2}^2)\: \ell_1 \cdot k_1\: \ell_2 \cdot k_2$.

\subsection{Ladder Integrals: an All-Loops Conjecture}
\begin{figure}[tb]
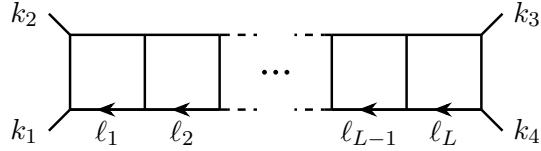

    \centering
    \Ladder
    \caption{%
      \label{fig:all_loop_ladder}%
      All-loop ladder integral.}
\end{figure}

Assuming that all integrals with ladder topologies have 
singularities that are at worst logarithmic in the IR (as is plausible), 
the reasoning in the previous section can be extended directly to 
all loop orders. In order to do so, we generalize \cref{eq:ladder_building blocks_2loop},
\begin{equation}%
  \label{eq:ladder_building blocks}
  \begin{gathered}   
    \FlowSymb_1 = \ell_1 \! \cdot \!v_2\,, \quad 
    \FlowSymb_2 = (\ell_1 -k_{12}) \! \cdot \!v_1\, ,\quad 
    \FlowSymb_3 = (\ell_L-k_{12}) \! \cdot \!v_2\,, \quad 
    \FlowSymb_4 = \ell_L \! \cdot \!(v_2-v_3)\, , 
    \\ 
    \FlowSymb_{12} = \ell_1 \! \cdot \!v_3\, , \quad
    \FlowSymb_{34} = (\ell_L- k_{123}) \! \cdot \!(v_1 - v_2)\, .
  \end{gathered}
\end{equation}
Using these variables and defining the $L$-loop momenta as in \cref{fig:all_loop_ladder}, 
our conjecture for the generators of the truncated ideal of 
locally finite numerators takes the remarkably simple form,
\begin{equation}\label{eq:ladder_conjecture}
    \begin{aligned}
        &\text{rank two}: &&\quad 
        \FlowSymb_{12} \:\FlowSymb_{34}\, , \quad 
       \LLort{1,}{L}\, ,
        \\
        &\text{rank three}: &&\quad 
        \FlowSymb_{1} \:\FlowSymb_{2} \:\FlowSymb_{34}\, , \quad 
        \FlowSymb_{12} \:\FlowSymb_{3} \:\FlowSymb_{4}\, , \quad
        \LLort{1,}{i \neq L} \:\FlowSymb_{34}\, , \quad
        \LLort{L,}{i \neq 1} \: \FlowSymb_{12}\, ,
        \\
        &\text{rank four}: &&\quad 
        \FlowSymb_{1} \:\FlowSymb_{2} \:\FlowSymb_{3} \:\FlowSymb_{4}\, , \quad 
        \LLort{1,}{i \neq L} \:\FlowSymb_{3} \:\FlowSymb_{4}\, , \quad
        \LLort{L,}{i \neq 1} \:\FlowSymb_{1} \:\FlowSymb_{2}\, , \quad
        \LLort{1,}{i \neq L} \LLort{L,}{j \neq 1} \, .
        \\
    \end{aligned}
\end{equation}
We thus have two conjectured generators at rank two, 
$2L$ generators at rank three,
$L^2$ generators at rank four, with no additional generators beyond rank four.
The larger number of generators is simply due to having more loop momenta 
at our 
disposal to build some of the \(\LLort{i}{j}\) combinations. The role of the 
scattering-plane variables $\FlowSymb$ is  unchanged.

As to evanescent generators, we see that (see \cref{sec:evanescent}) every Gram 
determinant involving at least two different $\Lort{i}$ will automatically vanish 
as $D\to4$. 
It follows that the truncated ideal of evanescent numerators is the intersection 
of the locally finite one of \cref{eq:ladder_conjecture} with 
the one generated by the two-by-two Gram determinants,
\begin{equation}
     \Gram{\Lort{i_1} & \Lort{i_2}}{\Lort{j_1} & \Lort{j_2}}  = \LLort{i_1}{j_1}\:\LLort{i_2}{j_2} - \LLort{i_1}{j_2}\:\LLort{i_2}{j_1}\, .
\end{equation}
We can write out the intersection explicitly, finding that it
is generated by the set,
\begin{equation}\label{eq:ladder_conjecture_evanescent}
    \begin{aligned}
        &\text{rank four}: \quad 
        &&\LLort{1}{i}\:\LLort{L}{j} - \LLort{1}{j}\:\LLort{L}{i}\, , \quad 
        \LLort{1}{L}\:\LLort{k_1}{h_1} - \LLort{1}{h_1}\:\LLort{k_1}{L}\, ,
        \\
        &\text{rank five}: \quad 
        &&(\LLort{1}{k_1}\:\LLort{k_3}{k_2} - \LLort{1}{k_2}\:\LLort{k_3}{k_1}) \:\FlowSymb_{34}\, , \quad 
        (\LLort{L}{h_1}\:\LLort{h_3}{h_2} - \LLort{L}{h_2}\:\LLort{h_3}{h_1})\: \FlowSymb_{12}\, ,
        \\
        &\text{rank six}: \quad 
        &&(\LLort{1}{k_1}\:\LLort{k_3}{k_2} - \LLort{1}{k_2}\:\LLort{k_3}{k_1}) \:\FlowSymb_{3} \:\FlowSymb_{4}\, , \quad 
        (\LLort{L}{h_1}\:\LLort{h_3}{h_2} - \LLort{L}{h_2}\:\LLort{h_3}{h_1}) \: \FlowSymb_{1} \:\FlowSymb_{2}\, , \\
        & &&  (\LLort{l_1}{l_3}\:\LLort{l_2}{l_4} - \LLort{l_1}{l_4}\:\LLort{l_2}{l_3}) \FlowSymb_{12} \: \FlowSymb_{34}\, ,
        \\
        &\text{rank seven}: \quad 
        &&(\LLort{l_1}{l_3}\:\LLort{l_2}{l_4} - \LLort{l_1}{l_4}\:\LLort{l_2}{l_3})  \: \FlowSymb_{12}\:\FlowSymb_{3} \:\FlowSymb_{4}\, , \quad 
        (\LLort{l_1}{l_3}\:\LLort{l_2}{l_4} - \LLort{l_1}{l_4}\:\LLort{l_2}{l_3})  \: \FlowSymb_{1} \:\FlowSymb_{2} \: \FlowSymb_{34} \, ,
        \\
        &\text{rank eight}: \quad 
        &&(\LLort{l_1}{l_3}\:\LLort{l_2}{l_4} - \LLort{l_1}{l_4}\:\LLort{l_2}{l_3})  \: \FlowSymb_{1} \: \FlowSymb_{2} \:\FlowSymb_{3} \:\FlowSymb_{4}\, ,
    \end{aligned}
\end{equation}
where the indices are defined as,
\begin{equation}
  i,j = 1, \dots, L, \quad
  k_r = 1, \dots, L-1,\quad
  h_r = 2, \dots, L, \quad
  l_r = 2, \dots, L-1 \,,
\end{equation}
in order to avoid double-counting of lower-rank generators and are assumed to take values for which the corresponding generator is non-zero, \textit{e.g.} the choice $l_1=l_2=l_3=l_4=2$ is forbidden because most generators at 
ranks six, seven, and eight would vanish identically.
The evanescent generators above can be written in terms of the
locally finite generators in \cref{eq:ladder_conjecture}.
From \cref{eq:ladder_conjecture_evanescent} we find the following numbers of generators at each rank:
\begin{equation}\label{eq:ladder_conjecture_evanescent_counts}
    \begin{aligned}
        &\text{rank four}: \quad 
        &&(3L^2-9L+8)/2\, ,
        \\
        &\text{rank five}: \quad 
        &&(L-2)(L^2-4L+5)\, ,
        \\
        &\text{rank six}: \quad 
        && (L-2) (L^3 - 9 L + 16)/8\, , 
        \\
        &\text{rank seven}: \quad 
        && (L - 2) (L - 3) (L^2 - 5 L + 8)/4 \, ,
        \\
        &\text{rank eight}: \quad 
        && (L - 2) (L - 3) (L^2 - 5 L + 8)/8 \, .
    \end{aligned}
\end{equation}
We deduce that the rank-four generators are present for any number of loops, 
while those at ranks five and six require a minimum of three loops, and
those at ranks seven and eight first appear at four loops. 
All locally finite and evanescent generators in
\cref{eq:ladder_conjecture,eq:ladder_conjecture_evanescent} are UV finite. 

We have verified the validity of this conjecture at two, three and four
loops by explicit computation of the ideals.

\subsection{Non-Planar Double Box}%
\label{sec:nonplanar_dbox}
We next consider the non-planar double box integral.
There are two important differences between the planar and non-planar double 
box integrals. In particular the non-planar one:
\begin{enumerate}
    \item lacks a subtraction-free form (see \cref{sec:landau_solve}) for the $\mathcal{F}$ polynomial;
    \item has power-like soft divergences: regions of the loop integration where the integrand scales as $\mathrm{d}\lambda \; \lambda^{-\alpha}$ with $\alpha>1$.
\end{enumerate}
For these reasons we believe the non-planar box to be an instructive example.
We repeat the procedure outlined in the previous section, 
highlighting the role of power-like divergeneces along the way.

The loop momenta are parametrized as shown in \cref{fig:nonplanar_dbox_graph} with all external momenta outgoing and we label the edges from 1 to 7 so that their momenta read, in increasing order of labels,
\begin{equation}
  \ell_1,\quad
  \ell_1 - k_1,\quad
  \ell_1 - k_{12},\quad
  \ell_2,\quad
  \ell_2 - k_{123},\quad
  \ell_2 - \ell_1 - k_3,\quad
  \ell_1 - \ell_2.
\end{equation}
As before, we start by imposing UV-convergence constraints and proceed to the IR analysis afterwards.

\subsubsection{UV divergences}
\begin{figure}[tb]
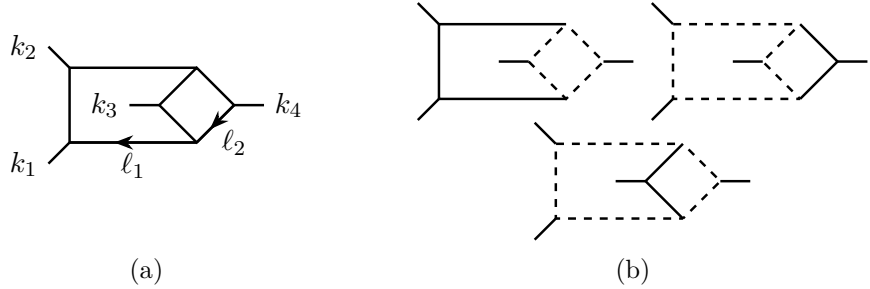

  \centering
  \newcommand{\SubfigureB}{
    \(
    \setlength{\extrarowheight}{5ex}
    \begin{array}[t]{cc}
      \NplDBoxBeforeA &
      \NplDBoxBeforeC \\
      \multicolumn{2}{c}{\NplDBoxBeforeB} 
    \end{array}
    \)
  }
  \subfloat[]{
    \label{fig:nonplanar_dbox_graph}
    \vphantom{\SubfigureB}\NplDBox
  }
  \qquad
  \subfloat[]{
    \label{fig:nonplanar_dbox_reductions}
    \SubfigureB
  }
  \caption{The non-planar double box graph (a) and its sub-integrations (b).}
  \label{fig:nonplanar_dbox}
\end{figure}

The maximal allowed numerator rank is five and there are three distinct 
sub-integrations corresponding to the dashed lines in \cref{fig:nonplanar_dbox_reductions}.
These give rise to the following constraints on the numerator:
\begin{equation}
 \lim_{\rho \to \infty} 
\begin{cases}
 \rho^{-4} \: \mathcal{N}\left( \ell_1 = \mathrm{const} , \;\ell_2 = \rho \ell \right) = 0\,, \\
 \rho^{-6} \: \mathcal{N}\left( \ell_1 = \rho \ell, \; \ell_2 = \mathrm{const} \right) = 0\,, \\
 \rho^{-6} \: \mathcal{N}\left( \ell_1 = \rho \ell, \;\ell_2 = \mathrm{const} + \rho \ell \right) = 0\, . 
\end{cases}
\end{equation}
Overall, a UV-finite numerator for the non-planar double box should be at 
most of rank five, with maximum degrees of five and three in $\ell_1$ and
$\ell_2$ respectively. 

\subsubsection{IR divergences}

As before, the relevant reduced diagrams correspond to the zeros of the first Symanzik polynomial:
\begin{equation}
  \mathcal{U} = \left( \alpha_1 + \alpha_2 + \alpha_3 \right) \left( \alpha_4 + \alpha_5 + \alpha_6 + \alpha_7 \right) + \left( \alpha_4 + \alpha_5 \right) \left( \alpha_6 + \alpha_7 \right),
\end{equation}
and reproduce the sub-integrations of \cref{fig:nonplanar_dbox_reductions}.
To solve the parameter-space Landau equations for the full diagram, we compute the second Symanzik polynomial:
\begin{equation}
  \mathcal{F} = s \left[ \alpha_1 \alpha_3 \left( \alpha_4 + \alpha_5 + \alpha_6 + \alpha_7 \right) + \alpha_1 \alpha_5 \alpha_6 + \alpha_3 \alpha_4 \alpha_7 \right]
  + t \, \alpha_2 \alpha_5 \alpha_7
  - (s + t) \, \alpha_2 \alpha_4 \alpha_6\,.
\end{equation}
Unlike the planar case, \(\mathcal{F}\) is no longer a sum
of positive terms for appropriate signs of the Mandelstam invariants.
Therefore, we consider instead the most general linear combination of the Landau equations~\eqref{eq:landau_par_lin_comb},
\begin{equation}
  \begin{aligned}
    \sum_{e=1}^E w_e \alpha_e \frac{\partial}{\partial \alpha_e} \mathcal{F} &=
    s \, (w_1 + w_3 + w_4) \, \alpha_1 \alpha_3 \alpha_4 + s \, (w_1 + w_3 + w_5) \, \alpha_1 \alpha_3 \alpha_5 \\
    &+ s \, (w_1 + w_3 + w_6) \, \alpha_1 \alpha_3 \alpha_6 + s \, (w_1 + w_3 + w_7) \, \alpha_1 \alpha_3 \alpha_7 \\
    &+ s \, (w_1 + w_5 + w_6) \, \alpha_1 \alpha_5 \alpha_6 + s \, (w_3 + w_4 + w_7) \, \alpha_3 \alpha_4 \alpha_7 \\
    &+ t \, (w_2 + w_5 + w_7) \, \alpha_2 \alpha_5 \alpha_7 - (s+t) \, (w_2 + w_4 + w_6) \, \alpha_2 \alpha_4 \alpha_6\,,
  \end{aligned}
\end{equation}
and require that each coefficient is strictly positive when \(s, t > 0\).
It is straightforward to check that the resulting system of linear inequalities is feasible: one possible solution is,
\begin{equation}
  \vec{w} = (0, -1, 1, 0, 2, 0, 0)\,.
\end{equation}
This means that a subtraction-free linear combination of the Landau equations exists, therefore we can find their solutions simply by setting \(\mathcal{F}\) to zero term by term as usual; in other words, no divergences due to cancellations in~\(\mathcal{F}\) can occur (see also ref.~\cite[Appendix~A]{Gardi:2022khw}). 
For the reduced diagrams, \(\mathcal{F}\) is subtraction-free itself, so their analysis does not pose any difficulty.

Solutions of the Landau equations correspond to double- and single-collinear 
IR divergences, all of which are logarithmic.
However, if we take intersections of these solutions we find additional 
subdivergences, two of which are power-like.  These are the double-soft 
configurations \((\ell_1 = 0, \ell_2 = 0)\) and 
\((\ell_1 = k_1 + k_2, \ell_2 = -k_4)\), in accordance with ref.~\cite{Anastasiou:2018rib}.
Therefore, unlike the planar case, cancellation of collinear 
configurations alone is insufficient.
In principle, one has to expand the UV-finite ansatz to a certain sub-leading order in the vicinity of the power-like configurations, and require that these divergent contributions vanish.
In practice however, we find that these additional constraints are satisfied automatically as long as all collinear divergences are canceled.
We report the results of our analysis in \cref{tab:npl_dbox_count}.
\begingroup
\renewcommand{\arraystretch}{\mysize}
\begin{table}[tb]
  \caption{\label{tab:npl_dbox_count}%
    Results of the non-planar double box numerator analysis.}
  \centering
  \begin{tabular}{lccccc}
    rank & 1 & 2 & 3 & 4 & 5 \\
    \midrule
    \# finite integrals & 0 & 0 & 9 & 65 & 230 \\
    \# finite generators & 0 & 0 & 9 & 8 & 0 \\
    \# evanescent integrals & 0 & 0 & 0 & 1 & 7 \\
    \# evanescent generators & 0 & 0 & 0 & 1 & 0 \\
  \end{tabular}
\end{table}
\endgroup

\subsubsection{Finding a Basis of Generators}

By analogy with the planar double box we can define a set of variables,
\begin{equation}%
  \label{eq:npl_doublebox_building blocks}
  \begin{gathered}   
    \FlowSymb_1 = \ell_1 \! \cdot \!v_2\,, \quad 
    \FlowSymb_2 = (\ell_1 -k_{12}) \! \cdot \!v_1\, ,\quad 
    \FlowSymb_3 = (\ell_1 - \ell_2) \! \cdot \!v_1 \,, \quad 
    \FlowSymb_3' = (\ell_1 - \ell_2) \! \cdot \!v_2 \,, 
    \\ 
    \FlowSymb_4 =  \ell_2 \! \cdot \!(v_1 - v_2)\, , \quad 
    \FlowSymb_4' =  \ell_2 \! \cdot \!(v_2 - v_3)\, , \quad 
    \FlowSymb_{12} = \ell_1 \! \cdot \!v_3  \, , 
  \end{gathered}
\end{equation}
which vanish on appropriate collinear and soft limits. In particular each $\FlowSymb_S$ (and $\FlowSymb_S'$) vanishes on the collinear configurations 
involving the particles in $S$, and also vanishes on the soft limits involving the combinations of loop momenta in its definition. In terms of these quantities we find the generators
\begin{equation} \label{eq:npl_double_box_generators_results}
    \begin{aligned}
        &\text{rank three}: \quad &&
        \FlowSymb_{12} \: \FlowSymb_{3} \:\FlowSymb_{4}\, , \quad 
        \FlowSymb_{12} \:\FlowSymb_{3}' \: \FlowSymb_{4}\, , \quad 
        \FlowSymb_{12} \:\FlowSymb_{3}  \:\FlowSymb_{4}'\, , \quad 
        \FlowSymb_{12} \:\FlowSymb_{3}' \:\FlowSymb_{4}'\, , \quad
        \FlowSymb_{12} \:( \LLort{1}{2} - \LLort{2}{2} ) \, , \\
        & && 
        \FlowSymb_{3} \: \LLort{1}{2} \, , \quad 
        \FlowSymb_{3}'\: \LLort{1}{2} \, , \quad 
        \FlowSymb_{4} \: ( \LLort{1}{1} - \LLort{1}{2} )  \, , \quad 
        \FlowSymb_{4}'\: ( \LLort{1}{1} - \LLort{1}{2} )  \, ,  \\[8pt]
        &\text{rank four}: \quad &&
        \FlowSymb_{1} \: \FlowSymb_{2} \: \FlowSymb_{3} \:\FlowSymb_{4}\, , \quad 
        \FlowSymb_{1} \: \FlowSymb_{2} \:\FlowSymb_{3}' \: \FlowSymb_{4}\, , \quad 
        \FlowSymb_{1} \: \FlowSymb_{2} \:\FlowSymb_{3}  \:\FlowSymb_{4}'\, , \quad 
        \FlowSymb_{1} \: \FlowSymb_{2} \:\FlowSymb_{3}' \:\FlowSymb_{4}'\, , \\
        & && 
        \beta_1 \beta_2 \: ( \LLort{1}{2} - \LLort{2}{2} )\, , \quad 
        \LLort{1}{2} \: ( \LLort{1}{1} - \LLort{1}{2} ) \, , \quad 
        \LLort{1}{2} \: ( \LLort{1}{2} - \LLort{2}{2} ) \, , \quad 
        \LLort{1}{1} \LLort{2}{2}- \LLort{1}{2}^2 \, .
    \end{aligned}
\end{equation}
Just as for the planar double box, we have only three independent 
external and two loop momenta,
so we expect a single evanascent generator.
This is indeed the case: it is 
$\GramO{\Lort{1} & \Lort{2}} = \LLort{1}{1} \: \LLort{2}{2}- \LLort{1}{2}^2 $.

\subsection{Two-Loop Beetle} \label{sec:beetle}
\begin{figure}[tb]
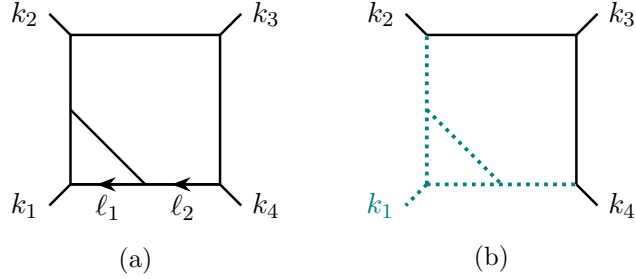

  \centering
  \subfloat[]{
    \label{fig:beetle_graph}
    \Beetle
  }
  \quad
  \subfloat[]{
    \label{fig:beetle_graph_collinear}
    \BeetleCollinearPowerlike
  }
  \caption{The beetle graph (a) and a collinear configuration leading to a power-like divergence (b). In (b) teal dotted edges are collinear to $k_1$.}
  \label{fig:beetle}
\end{figure}

In this section we briefly treat the \enquote{beetle} integral of 
\cref{fig:beetle_graph}.  We will not show the whole procedure here, but simply 
provide results for the locally finite and evanescent truncated ideals.  

This integral is of interest to us because it contains power-like double-collinear (see \cref{fig:beetle_graph_collinear}) and soft-collinear divergences, which did 
not appear in any of cases treated above. 
The presence of power-like divergences involving collinear configurations requires us to take into account \cref{eq:extra_relations} for the first time in this 
article. In practice, as we will see shortly, this introduces a dependence on the 
Mandelstam variables $s_{ij}$ in the set of generators for the ideal of locally
finite numerators.
To describe our results, we can define the set of rank-one monomials,
\begin{equation} \label{eq:beetle_building blocks}
    \begin{split}   
        &\FlowSymb_1 = \ell_1 \! \cdot \!v_2\,, \quad 
        \FlowSymb_{12} = \ell_1 \! \cdot \!v_3\,, \quad 
        \FlowSymbX_1 = \ell_2 \! \cdot \!v_2\,, \quad 
        \FlowSymbX_3 = (\ell_2 -k_{12}) \! \cdot \!v_2\,,
        \\
        &
        \FlowSymbX_{12} = \ell_2 \! \cdot \!v_3\,, \quad 
        \FlowSymbX_{14} = \ell_2 \! \cdot (v_2-v_3)\,, \quad
        \FlowSymbX_{34} = \ell_2 \! \cdot (v_1-v_2)\,, \quad
        \FlowSymbX_{23} = (\ell_2-k_1) \! \cdot \!v_1\,, \quad 
    \end{split}
\end{equation}
so that the $\FlowSymb_S$ (respectively the $\FlowSymbX_S$) vanish on collinear configurations of $\ell_1$ (respectively $\ell_2$) involving the 
external momenta in $S$. In terms of these variables\footnote{The 
variables in \cref{eq:beetle_building blocks} are not all linearly independent. We 
express our results in terms of an overcomplete set for the sake of clarity.}, we 
can express the set of generators as follows,
\begin{equation}
    \begin{aligned}
        &\text{rank three}: \quad &&
        \FlowSymbX_{12} \: \FlowSymbX_{34} \: ( s_{12} \FlowSymb_{1} +  s_{13} \FlowSymb_{12}) \, , \quad 
        \FlowSymbX_{14} \: \FlowSymbX_{23} \: ( s_{12} \FlowSymb_{1} +  s_{13} \FlowSymb_{12})  \, ,  
        \\
        & && 
        \FlowSymb_{1} \: \LLort{2}{2} \, , \quad 
        \FlowSymb_{12} \: \LLort{2}{2} \, , \quad 
        \FlowSymbX_{1} \: \LLort{1}{2} \, , \quad 
        \FlowSymbX_{12} \: \LLort{1}{2} \, ,         
        \\[8pt]
        &\text{rank four}: \quad &&
        \FlowSymb_{1} \:\FlowSymbX_{12} \: \FlowSymbX_{34} \: ( s_{12} \FlowSymbX_{1} +  s_{13} \FlowSymbX_{12})\, , \quad 
       \FlowSymb_{1} \:  \FlowSymbX_{14} \: \FlowSymbX_{23} \: ( s_{12} \FlowSymbX_{1} +  s_{13} \FlowSymbX_{12})\, , \quad         
        \\
        & && 
        \FlowSymb_{12} \:\FlowSymbX_{12}^2 \: \FlowSymbX_{34} \, , \quad 
         \FlowSymb_{12} \:\FlowSymbX_{12} \: \FlowSymbX_{23} \: \FlowSymbX_{14} \, ,        
         \\
        & && 
        \FlowSymb_{1} \: \FlowSymbX_{23} \:  [ s_{23} \FlowSymbX_{1}^2 +  (\FlowSymbX_{1} + \FlowSymbX_{34})( s_{12} \FlowSymbX_{1} +  s_{13} \FlowSymbX_{12})]\, , \quad 
        \LLort{1}{2}  \: \LLort{2}{2} \, .  
    \end{aligned}
\end{equation}
The polynomial $\beta_1 s_{12} + \beta_{12} s_{13}$ and its equivalent with $\beta \to \gamma$ are related to standard scalar products of loop and external momenta,
\begin{equation}
     \beta_1 s_{12} + \beta_{12} s_{13} = 2 k_1 \cdot \ell_1 
     \quad \text{and} \quad
     \gamma_1 s_{12} + \gamma_{12} s_{13} = 2 k_1 \cdot \ell_2 \, .
\end{equation}

Finally we find that, within the UV power counting, there are no evanescent numerators for the beetle integral. 

\subsection{Two-Loop Four-Gluon Helicity Amplitudes}
In this section, we present a simple but concrete application
of the concepts and results discussed in previous sections.
We focus on the leading-color amplitudes for two-loop scattering of 
four gluons in pure Yang--Mills theory. 
Very compact expressions for the all-plus amplitude have been known for a long time~\cite{Bern:2000dn} and expressions for all other helicity configurations were first computed in refs.~\cite{Bern:2001df} and~\cite{Glover:2001af}.
Our goal here is not to obtain the most compact results, but rather to give a hint of how organising scattering amplitudes according to their IR structure 
can offer a simpler representation. 
For our example, we make use of the rank-two locally finite numerators of \cref{eq:double_box_generators_results},
\begin{equation}\label{eq:double_box_rank2_numerators}
    f_1(k_1,k_2,k_3,k_4) = \beta_{12} \beta_{34}\, , \quad  f_2(k_1,k_2,k_3,k_4) = \LLort{1}{2}\,,
\end{equation}
to write the bare four-gluon helicity amplitudes in a basis of 13 master 
integrals as follows,
\begin{gather} 
    \mathcal{A}^{0\to gggg}(\bm \lambda) = 
   \Bigg[ 
    r_1^{\bm \lambda} \resizebox{0.1\textwidth}{!}{\doublebox} [f_1(k_1,k_2,k_3,k_4)] + 
    r_2^{\bm \lambda} \resizebox{0.1\textwidth}{!}{\doublebox} [f_2(k_1,k_2,k_3,k_4)] + 
    \notag \\
    r_3^{\bm \lambda} \left.\resizebox{0.06\textwidth}{!}{\doubleboxcrossed} [f_1(k_4,k_1,k_2,k_3)]\right. +  
    r_4^{\bm \lambda} \left.\resizebox{0.06\textwidth}{!}{\doubleboxcrossed} [f_2(k_4,k_1,k_2,k_3)]\right. +
    \notag \\
    r_5^{\bm \lambda} \resizebox{0.1\textwidth}{!}{\trianglebox} +
    r_6^{\bm \lambda} \left.\resizebox{0.065\textwidth}{!}{\triangleboxcrossed}\,\right. +
    r_7^{\bm \lambda} \resizebox{0.065\textwidth}{!}{\slashbox} +
    r_8^{\bm \lambda} \resizebox{0.065\textwidth}{!}{\slashboxsix} +
    r_9^{\bm \lambda} \left.\resizebox{0.065\textwidth}{!}{\slashboxsixcrossed} \right. + 
    \notag \\
    r_{10}^{\bm \lambda} \resizebox{0.1\textwidth}{!}{\doubletriangle} +
    r_{11}^{\bm \lambda} \left.\resizebox{0.06\textwidth}{!}{\doubletrianglecrossed}\,\right. +
    r_{12}^{\bm \lambda} \resizebox{0.1\textwidth}{!}{\teepee} +
    r_{13}^{\bm \lambda} \left.\resizebox{0.07\textwidth}{!}{\teepeecrossed}\,\right.
    \Bigg] 
    \Phi(\bm \lambda)
    \, .
\label{eq:master_planar_basis}
\end{gather}
Here, $\mathcal{A}^{0\to gggg}(\bm \lambda)$ is the coefficient of the color 
factor $N_c^2\: \mathrm{Tr}[T^{a_1}T^{a_2}T^{a_3}T^{a_4}]$ for the helicity configuration $\bm \lambda = \{\lambda_1,\lambda_2,\lambda_3,\lambda_4 \}$
($N_c$ is the dimension of the Yang--Mills group and $a_i$ is the color index of 
the $i$--th gluon). The coefficients $r_i^{\bm \lambda}$ are rational functions of the space-time dimensions $D$ and of the independent Mandelstam variables $s_{12},s_{23}$. Finally, $\Phi(\bm \lambda)$ is an overall spinor factor.

The first four integrals in \cref{eq:master_planar_basis} are locally finite, and contain the whole dependence on the double-box topology (the one with the 
largest number of propagators). 
The numerators of \cref{eq:double_box_rank2_numerators} correspond to linear combinations of the chiral numerators presented in ref.~\cite{Caron-Huot:2012awx}.
The remaining nine integrals are chosen to match as closely as possible the IR singularities obtained from the Landau equations. Because our analysis is for the moment limited to IR-finite integrals we do not have a systematic way to select IR-divergent integrals. We postpone an investigation of this aspect to future work.

With the choice of master integrals adopted in \cref{eq:master_planar_basis}, we observe that all rational coefficients $r_i^{\bm \lambda}$ are regular in the limit $\epsilon \to 0$, implying that all divergences are contained in the integrals themselves.

The helicity configurations which vanish at tree-level have an overall factor of 
$\epsilon$ so that, so long as we are interested in the amplitudes only up to $\Ord(\epsilon^0)$, we can safely drop all locally finite integrals. 
We then find, 
\def\ratcoeff{\Tilde r}
\begin{gather} 
    \mathcal{A}^{0\to gggg}(+,+,+,+) = \epsilon \: C(\epsilon) \left[ 
    \ratcoeff_5 \resizebox{0.1\textwidth}{!}{\trianglebox} +
    \ratcoeff_6 \left.\resizebox{0.065\textwidth}{!}{\triangleboxcrossed}\,\right. +
    \ratcoeff_7 \resizebox{0.065\textwidth}{!}{\slashbox} +
    \ratcoeff_8 \resizebox{0.065\textwidth}{!}{\slashboxsix} +
    \ratcoeff_9 \left.\resizebox{0.065\textwidth}{!}{\slashboxsixcrossed} \right.+ 
    \right.
    \notag \\
    \left.
    \ratcoeff_{10} \resizebox{0.1\textwidth}{!}{\doubletriangle} +
    \ratcoeff_{11} \left.\resizebox{0.06\textwidth}{!}{\doubletrianglecrossed}\,\right. +
    \ratcoeff_{12} \resizebox{0.1\textwidth}{!}{\teepee} +
    \ratcoeff_{13} \left.\resizebox{0.07\textwidth}{!}{\teepeecrossed}\,\right.
    \right]\Phi(+,+,+,+)\,,
    \label{eq:exact_gggg_++++}
\end{gather}
where we have defined the rational coefficients $\ratcoeff_i$ 
to make the overall factor of $\epsilon$ explicit,
and where the $\epsilon$-dependent coefficient $C(\epsilon)$ approaches a finite constant as $\epsilon \to 0$. We collect the explicit expressions for the 
coefficients in \cref{sec:explicit_coefficients}. 
A similar expression holds for the single-minus amplitude.
Our choice of master integrals makes \cref{eq:exact_gggg_++++} free of contributions related to the double box, the most complicated topology for this process.

\section{Conclusions}
\label{sec:conclusions}

In this article, we presented a systematic approach to finding locally finite
integrands for Feynman integrals.  Such integrands yield integrals which are
UV and IR finite, and which are integrable everywhere in loop-momentum space.
They can be evaluated in four dimensions.  The locally IR-finite integrands
form an ideal, which we can truncate to a finite-dimensional space of
UV-finite integrands.  We showed how to write the generators of the ideal
in a compact form using either dual vectors or Gram determinants.  

We also presented the class of evanescent integrands, a subset of
locally finite ones.  These integrands give rise to integrals which are of 
$\Ord(\eps)$ in the dimensional regulator, and so will vanish in the four-dimensional limit.  
They can give rise to new identities between
Feynman integrals.  We also briefly discussed evanescently finite
integrands.  These give rise to finite integrals whose finiteness
arises from cancellation of a UV or IR divergence with a factor of $\eps$ arising from the integration.  
Such integrands are special to dimensional regularization, but may play a role in providing expressions 
for rational terms beyond one loop.  We leave a more thorough investigation of such integrands to future work.  

We presented several explicit examples at two loops: the planar and non-planar double boxes, as well as the integral of \cref{fig:beetle}.  We also presented
a conjecture for the locally finite and evanescent integrands for all ladder integrals.  We verified the conjecture at three and four loops. 
Both classes of integrands will be pruned through use of integration-by-parts~(IBP) identities~\cite{Tkachov:1981wb,Chetyrkin:1981qh}, though ideally these would be applied in a way that respects finiteness or evanescence.  We believe a compatible approach is possible and an interesting subject for investigation.

The isolation of locally finite and evanescent integrands represents the first step in a program of reorganizing integrands in classes of uniform UV and IR divergence.  A next step, for example, could be isolating all integrals with $1/\eps$ divergences of IR origin.  We illustrated the promise of such a reorganization by showing that the expressions for planar two-loop four-gluon amplitudes
simplify considerably when we make use of finite integrals as master integrals.

The interplay
between finite or evanescent integrands and
the class of so-called local integrands~\cite{Arkani-Hamed:2010pyv}
also offers an interesting subject for investigation.
The local integrands allow one to express
scattering amplitudes in a manifestly local form. In addition, they  give rise
to integrals with especially simple analytic properties.
The integrals fulfil simple differential equations,
and can be expressed as pure
combinations of functions of uniform transcendental weight~\cite{Henn:2013pwa}.
The algorithms presented in this paper, in combination
with a suitable set of IBP identities as described above, 
offer an opportunity to investigate these classes of integrals from a 
more general perspective.

\begin{acknowledgements}
We thank Michael Borinsky, Leonardo de la Cruz, Andreas von Manteuffel,
Ben Page, Stefan Weinzierl, and Yang Zhang
for helpful discussions.

This work was supported in part by the Excellence
Cluster ORIGINS funded by the Deutsche Forschungsgemeinschaft (DFG, German Research Foundation) under
Germany’s Excellence Strategy – EXC--2094--390783311,
by the European Research
Council (ERC) under the European Union’s research and
innovation programme grant agreements ERC--AdG--885414 (`Ampl2Einstein')
and ERC--StG--949279 (`HighPHun'), and by the Royal Society under the Royal 
Society University Research Fellowship (URF/R1/191125).
\end{acknowledgements}

\appendix
\section{Gram Determinant Representation for Locally Finite Numerators}
In this section we present examples of Gram-determinant representations for 
the truncated ideals of locally finite numerators. 

\subsection{Double Box}%
\label{sec:doublebox_gram_ideal}
We found that \cref{eq:double_box_generators_results} is a complete basis of locally finite numerators for the double-box topology. An equivalent basis can be obtained in terms of Gram determinants and reads,

\begin{equation}\label{eq:double_box_generators_results_gram_representation}
    \begin{aligned}
        &\text{rank two}: \quad
        &&\Gram{\ell_1&1&2}{\ell_2&3&4}\, , \quad \Gram{\ell_1&1&2&3}{\ell_2&1&2&3}\, ,
        \\
        &\text{rank three}: \quad
        &&
        \GramO{\ell_2 & 3 & 4}\,
        \Gram{\ell_1 & 1 & 2}{1 & 2 & 4}
        \,, \quad
        \GramO{\ell_1 & 1 & 2}\,
        \Gram{\ell_2 & 3 & 4}{1 & 2 & 4}
        \,,
        \\
        & &&
        \left( \ell_2+k_4 \right)^2\,
        \Gram{\ell_1 & 1 & 2}{1 & 2 & 4}
        \,, \quad
        \left( \ell_1-k_1 \right)^2\,
        \Gram{\ell_2 & 3 & 4}{1 & 2 & 4}
        \\
        &\text{rank four}: \quad
        &&
        \left( \ell_2+k_4 \right)^2\,
        \GramO{\ell_1 & 1 & 2}
        \,, \quad
        \left( \ell_1-k_1 \right)^2\,
        \GramO{\ell_2 & 3 & 4}
        \,,
        \\
        & &&
        \left( \ell_1-k_1 \right)^2\,
        \left( \ell_2+k_4 \right)^2
        \,, \quad
        \GramO{\ell_1&\ell_2&1&2&3} \, .
    \end{aligned}
\end{equation}
The lone rank-four evanescent generator of \cref{eq:double_box_generators_results_evanescent} corresponds to the Gram determinant,
\begin{equation} \label{eq:2loop_evanescent_gram}
    \GramO{\ell_1&\ell_2&1&2&3}.
\end{equation}
The basis in \cref{eq:double_box_generators_results_gram_representation} is not unique; in general one can take linear combinations of the generators to construct 
different bases. For example, an alternative choice for the rank-two 
generators is,
\begin{equation}
 \Gram{\ell_1&1&2}{\ell_2&3&4}\, , \quad \Gram{\ell_1&1&2}{1&2&3}\Gram{\ell_2&3&4}{1&2&3} \,.
\end{equation}

\subsection{Non-Planar Double Box}%
\label{sec:npldoublebox_gram_ideal}
One possible choice of Gram-determinant representation for the locally finite basis of \cref{eq:npl_double_box_generators_results} is,
  \begin{equation}
      \begin{aligned}
        &\text{rank three}: 
        &&
        \Gram{\ell_1 & 1 & 2}{1 & 2 & 3}\,
        \Gram{\ell_2 & 4}{1 & 2}\,
        \Gram{\ell_1 - \ell_2 & 3}{1 & 2}
        \,, \quad
        \Gram{\ell_1 & 1 & 2}{1 & 2 & 3}\,
        \Gram{\ell_2 & 4}{2 & 3}\,
        \Gram{\ell_1 - \ell_2 & 3}{2 & 3}
        \,,
        \\
        & &&
        \Gram{\ell_1 & 1 & 2}{1 & 2 & 3}\,
        \Gram{\ell_2 & 4}{3 & 4}\,
        \Gram{\ell_1 - \ell_2 & 3}{3 & 4}
        \,, \quad
        \Gram{\ell_1 & 1 & 2}{1 & 2 & 3}\,
        \Gram{\ell_2 & 4}{4 & 1}\,
        \Gram{\ell_1 - \ell_2 & 3}{4 & 1}
        \,,
        \\
        & &&
        \Gram{\ell_2 & 4}{\ell_1 - \ell_2 & 3}\,
        \Gram{\ell_1 & 1 & 2}{1 & 2 & 3}
        \,, \quad 
        \Gram{\ell_2 & 4}{1 & 2}\,
        \Gram{\ell_1 & 1 & 2}{\ell_1 - \ell_2 & 3 & 1}
        \,,
        \\
        & &&
        \Gram{\ell_2 & 4}{2 & 3}\,
        \Gram{\ell_1 & 1 & 2}{\ell_1 - \ell_2 & 3 & 1}
        \,, \quad
        \Gram{\ell_1 - \ell_2 & 3}{1 & 2}\,
        \Gram{\ell_1 & 1 & 2}{\ell_2 & 4 & 1}
        \,,
        \\
        & &&
        \Gram{\ell_1 - \ell_2 & 3}{2 & 3}\,
        \Gram{\ell_1 & 1 & 2}{\ell_2 & 4 & 1}
        \,,
        \\
        &\text{rank four}:
        &&  
        \left( \ell_1-k_1 \right)^2\,
        \Gram{\ell_2 & 4}{1 & 2}\,
        \Gram{\ell_1 - \ell_2 & 3}{1 & 2}
        \,,\quad   
        \left( \ell_1-k_1 \right)^2\,
        \Gram{\ell_2 & 4}{2 & 3}\,
        \Gram{\ell_1 - \ell_2 & 3}{2 & 3}
        \,,
        \\
        & &&
        \left( \ell_1-k_1 \right)^2\,
        \Gram{\ell_2 & 4}{3 & 4}\,
        \Gram{\ell_1 - \ell_2 & 3}{3 & 4}
        \,,\quad   
        \left( \ell_1-k_1 \right)^2\,
        \Gram{\ell_2 & 4}{4 & 1}\,
        \Gram{\ell_1 - \ell_2 & 3}{4 & 1}
        \,,
        \\
        & &&
        \GramO{\ell_1 & 1 & 2}\,
        \Gram{\ell_2 & 4}{\ell_1 - \ell_2 & 3}
        \,,\quad
        \Gram{\ell_1 & 1 & 2}{\ell_2 & 1 & 2}\,
        \Gram{\ell_2 & 4}{\ell_1 - \ell_2 & 3}
        \,,
        \\
        & &&
        \left( \ell_1-k_1 \right)^2\,
           \Gram{\ell_2 & 4}{\ell_1 - \ell_2 & 3}
           \,, \quad 
           \GramO{\ell_1 & \ell_2 & 1 & 2 & 3}\,,
      \end{aligned}
  \end{equation}
while the lone evanescent generator is proportional to the same Gram determinant given in \cref{eq:2loop_evanescent_gram}.

\subsection{Two-Loop Beetle}%
\label{sec:beetle_gram_ideal}
As for the other topologies in this appendix, we provide the Gram determinant form of the locally finite ideal for the beetle integral defined in \cref{sec:beetle}. A possible choice of basis is,
\begin{equation}
  \begin{aligned}
    &\text{rank three}:
    &&
    \Gram{\ell_1 & 1}{1 & 2}\,
    \Gram{\ell_2 & 1 & 2}{\ell_2 & 3 & 4}
    \,, \quad
    \Gram{\ell_1 & 1}{1 & 2}\,
    \Gram{\ell_2 & 1 & 4}{\ell_2 - k_1 & 2 & 3}
    \,,
    \\
    & &&
    \GramO{\ell_2 & 1 & 2 & 4}\,
    \Gram{\ell_1 & 1}{1 & 2}
    \,, \quad
    \GramO{\ell_2 & 1 & 2 & 4}\,
    \Gram{\ell_1 & 1}{2 & 3}
    \,,
    \\
    & &&
    \Gram{\ell_2 & 1}{1 & 2}\,
    \Gram{\ell_2 & 1 & 2 & 3}{\ell_1 & 1 & 2 & 3}
    \,, \quad
    \Gram{\ell_2 & \ell_1 & 1 & 2}{\ell_2 & 1 & 2 & 3}
    \,,
    \\
    &\text{rank four}:
    &&
    \Gram{\ell_2 & 1}{1 & 2}\,
    \Gram{\ell_1 & 1}{2 & 3}\,
    \Gram{\ell_2 & 1 & 2}{1 & 2 & 3}\,
    \Gram{\ell_2 & 3 & 4}{1 & 2 & 3}
    \,,
    \\
    & &&
    \Gram{\ell_2 & 1}{1 & 2}\,
    \Gram{\ell_2 & 4}{1 & 2}\,
    \Gram{\ell_1 & 1}{2 & 3}\,
    \Gram{\ell_2 - k_1 & 2 & 3}{1 & 2 & 3}
    \,,
    \\
    & &&
    \Gram{\ell_2 & 1}{1 & 2}\,
    \Gram{\ell_1 & 1}{2 & 3}\,
    \Gram{\ell_2 & 1 & 4}{ \ell_2 - k_1 & 2 & 3}
    \,, \quad
    \Gram{\ell_1 & 1}{2 & 3}\,
    \Gram{\ell_2 & 3 & 4}{1 & 2 & 3}
    \Gram{\ell_2 & 1 & 2}{1 & 2 & 3}^2
    \,, \hspace*{-15mm}
    \\
    & &&
    \Gram{\ell_1 & 1}{2 & 3}\,
    \Gram{\ell_2 & 1 & 2}{1 & 2 & 3}\,
    \Gram{\ell_2 & 1 & 4}{\ell_2 - k_1 & 2 & 3}
    \,, \quad
    \GramO{\ell_2 & 1 & 2 & 3}\,
    \Gram{\ell_2 & 1 & 2 & 3}{\ell_1 & 1 & 2 & 3}
    \,.\hspace*{-10mm}
    \\
  \end{aligned}
\end{equation}

\section{Rational Coefficients of the All-Plus Amplitude}%
\label{sec:explicit_coefficients}
In this section we list the rational coefficients for the all-plus gluon amplitude of \cref{eq:exact_gggg_++++}. In terms of the variable $x = -s_{13}/s_{12}$ we find,
\def\colepsilon{{\color[rgb]{0.3,0.4,0.9}\epsilon}}
\begin{align}
         \Tilde{r}^+_{5} &= 
        \frac{1}{x}\!-\!\frac{1}{x^2}
        \!+\!\left(\frac{55}{6 x^2}\!-\!\frac{193}{6 x}\!-\!\frac{10}{x\!-\!1}\!-\!1\right) \colepsilon
        \!+\!\left(\!-\frac{92}{3 x^2}\!+\!\frac{218}{3 x}\!+\!\frac{82}{x\!-\!1}\!+\!\frac{49}{6}\right) \colepsilon ^2
       \nonumber  \\
        &\quad 
        \!+\!\left(\frac{151}{3 x^2}\!-\!\frac{205}{3 x}\!-\!\frac{513}{2 (x\!-\!1)}\!-\!25\right) \colepsilon ^3
        \, ,\nonumber \\ 
          \Tilde{r}^+_{6} &= 
        \!-\frac{1}{x^2}\!+\!x\!+\!\frac{3}{x}\!-\!3
        \!+\!\left(\!-10 x^3\!+\!6 x^2\!+\!\frac{55}{6 x^2}\!+\!\frac{191 x}{6}\!-\!\frac{9}{2 x}\!-\!\frac{65}{2}\right) \colepsilon
        \nonumber \\
        &\quad 
        \!+\!\left(82 x^3\!-\!\frac{1175 x^2}{6}\!-\!\frac{92}{3 x^2}\!+\!\frac{403 x}{3}\!+\!\frac{50}{x}\!-\!\frac{239}{6}\right) \colepsilon ^2  
        \nonumber \\
        &\quad 
        \!+\!\left(\!-\frac{513 x^3}{2}\!+\!\frac{1453 x^2}{2}\!+\!\frac{151}{3 x^2}\!-\!\frac{4295 x}{6}\!-\!\frac{133}{x}\!+\!\frac{657}{2}\right) \colepsilon ^3
        \, ,\nonumber \\ 
          \Tilde{r}^+_{7} &= 
        6\!-\!\frac{6}{x}
        \!+\!\left(3 x\!+\!\frac{55}{x}\!-\!55\right) \colepsilon
        \!+\!\left(\!-\frac{79 x}{2}\!-\!\frac{184}{x}\!+\!184\right) \colepsilon ^2
        \!+\!\left(\frac{291 x}{2}\!+\!\frac{302}{x}\!-\!302\right) \colepsilon ^3
        \, ,\nonumber \\ 
          \Tilde{r}^+_{8} &= 
        \!-\frac{1}{x^2}\!+\!\frac{2}{x}\!-\!1
        \!+\!\left(\frac{55}{6 x^2}\!-\!x\!-\!\frac{118}{3 x}\!+\!\frac{247}{6}\right) \colepsilon
        \!+\!\left(\!-\frac{92}{3 x^2}\!+\!\frac{49 x}{6}\!+\!\frac{87}{x}\!-\!\frac{253}{2}\right) \colepsilon ^2
        \nonumber \\
        &\quad 
        \!+\!\left(\frac{151}{3 x^2}\!-\!25 x\!-\!\frac{221}{3 x}\!+\!\frac{1205}{6}\right) \colepsilon ^3
        \, ,\nonumber \\ 
          \Tilde{r}^+_{9} &= 
        \!- \frac{1}{x^2}\!+\!\frac{2}{x}\!-\!1
        \!+\!\left(10 x^2\!+\!\frac{55}{6 x^2}\!+\!\frac{8}{3 x}\!-\!\frac{131}{6}\right) \colepsilon
        \!+\!\left(\!-62 x^2\!-\!\frac{92}{3 x^2}\!+\!\frac{213 x}{2}\!+\!\frac{107}{3 x}\!-\!\frac{99}{2}\right) \colepsilon ^2
        \nonumber \\
        &\quad 
        \!+\!\left(\frac{305 x^2}{2}\!+\!\frac{151}{3 x^2}\!-\!357 x\!-\!\frac{383}{3 x}\!+\!\frac{1691}{6}\right) \colepsilon ^3
        \, , \nonumber\\ 
          \Tilde{r}^+_{10} &= 
        \frac{1}{x^2}\!-\!\frac{1}{x}
        \!+\!\left(\!-\frac{55}{6 x^2}\!+\!\frac{193}{6 x}\!-\!\frac{5}{x\!-\!1}\!-\!41\right) \colepsilon
        \!+\!\left(\frac{92}{3 x^2}\!-\!\frac{218}{3 x}\!+\!\frac{11}{x\!-\!1}\!+\!\frac{767}{6}\right) \colepsilon ^2
       \nonumber \\
        &\quad 
        \!+\!\left(\!-\frac{151}{3 x^2}\!+\!\frac{x}{2}\!+\!\frac{231}{4 (x\!-\!1)}\!+\!\frac{205}{3 x}\!-\!\frac{117}{2}\right) \colepsilon ^3
        \, ,\nonumber \\ 
          \Tilde{r}^+_{11} &= 
        \frac{1}{x^2}\!-\!x\!-\!\frac{3}{x}\!+\!3
        \!+\!\left(\!-5 x^3\!-\!3 x^2\!-\!\frac{55}{6 x^2}\!+\!\frac{43 x}{6}\!+\!\frac{9}{2 x}\!+\!\frac{11}{2}\right) \colepsilon
        \nonumber \\
        &\quad 
        \!+\!\left(11 x^3\!+\!\frac{317 x^2}{6}\!+\!\frac{92}{3 x^2}\!-\!\frac{382 x}{3}\!-\!\frac{50}{x}\!+\!\frac{497}{6}\right) \colepsilon ^2
        \nonumber \\
        &\quad 
        \!+\!\left(\frac{231 x^3}{4}\!-\!\frac{853 x^2}{4}\!-\!\frac{151}{3 x^2}\!+\!\frac{3433 x}{12}\!+\!\frac{133}{x}\!-\!\frac{853}{4}\right) \colepsilon ^3
        \, ,\nonumber \\ 
          \Tilde{r}^+_{12} &= 
        \frac{6}{x^2}\!-\!\frac{12}{x}\!+\!6
        \!+\!\left(30 x^2\!-\!\frac{55}{x^2}\!+\!21 x\!+\!\frac{92}{x}\!-\!118\right) \colepsilon
        \!+\!\left(\!-126 x^2\!+\!\frac{184}{x^2}\!-\!19 x\!-\!\frac{221}{x}\!+\!368\right) \colepsilon ^2
        \nonumber \\
        &\quad 
        \!+\!\left(\frac{51 x^2}{2}\!-\!\frac{302}{x^2}\!-\!198 x\!+\!\frac{199}{x}\!-\!182\right) \colepsilon ^3
        \, , \nonumber \\ 
          \Tilde{r}^+_{13} &= 
        \frac{6}{x}\!-\!\frac{6}{x^2}
        \!+\!\left(\frac{55}{x^2}\!-\!30 x\!+\!\frac{30}{x\!-\!1}\!-\!\frac{73}{x}\!+\!129\right) \colepsilon
        \!+\!\left(\!-\frac{184}{x^2}\!+\!186 x\!-\!\frac{126}{x\!-\!1}\!+\!\frac{331}{x}\!-\!604\right) \colepsilon ^2
       \nonumber \\
        &\quad 
        \!+\!\left(\frac{302}{x^2}\!-\!\frac{915 x}{2}\!+\!\frac{51}{2 (x\!-\!1)}\!-\!\frac{707}{x}\!+\!\frac{1431}{2}\right) \colepsilon ^3 \, .
\end{align}
Terms of $\Ord(\epsilon^4)$ do not contribute to the finite part of the amplitude.
The overall coefficient reads,
\begin{equation}
    C(\epsilon) = \frac{16}{(-3 + 2 \epsilon)^2 (-1 + 2 \epsilon)^2 (-2 + 3 \epsilon) (-1 + 3 \epsilon)}\,.
\end{equation}

\bibliography{main.bib}

\end{document}